\documentclass[format=acmtog, review=false, screen=true]{acmart}
\usepackage{booktabs} 

\setcitestyle{square}

\usepackage{algorithm}  
\usepackage{algpseudocode}  
\usepackage{amsmath}  
\usepackage{sidecap}
\usepackage{wrapfig}

\usepackage{float}

\begin{document}
\title[P-Cloth]{P-Cloth: Interactive Complex Cloth Simulation on Multi-GPU Systems using Dynamic Matrix Assembly and Pipelined Implicit Integrators}


\renewcommand{\shortauthors}{papers\_108}

\author{Cheng Li}
\affiliation{%
  \institution{Zhejiang University}}
\email{licharmy@yahoo.com}

\author{Min Tang}
\affiliation{%
  \institution{Zhejiang University}}
\email{tang_m@zju.edu.cn}

\author{Ruofeng Tong}
\affiliation{%
  \institution{Zhejiang University}}
\email{trf@zju.edu.cn}

\author{Ming Cai}
\affiliation{%
  \institution{Zhejiang University}}
\email{cm@zju.edu.cn}

\author{Jieyi Zhao}
\affiliation{%
  \institution{University of Texas Health Science Center at Houston}}
\email{jieyi.zhao@uth.tmc.edu}

\author{Dinesh Manocha}
\affiliation{%
  \institution{University of Maryland at College Park}}
\email{dm@cs.umd.edu}

\renewcommand{\shortauthors}{C. Li et al.}


\begin{abstract}
We present a novel parallel algorithm for cloth simulation that exploits multiple GPUs for fast computation and the handling of very high resolution meshes. To accelerate implicit integration, we describe new parallel algorithms for sparse matrix-vector multiplication (SpMV)  and for dynamic matrix assembly on a multi-GPU workstation. Our algorithms use a novel work queue generation scheme for a fat-tree GPU interconnect topology. Furthermore, we present a novel collision handling scheme that uses spatial hashing for discrete and continuous collision detection along with a non-linear impact zone solver. Our parallel schemes can distribute the computation and storage overhead among multiple GPUs and enable us to perform almost interactive simulation on complex cloth meshes, which can hardly be handled on a single GPU due to memory limitations.
We have evaluated the performance with two multi-GPU workstations (with $4$ and $8$ GPUs, respectively)
on cloth meshes with $0.5-1.65M$ triangles. Our approach can reliably handle the collisions and generate vivid wrinkles and folds at $2-5$ fps, which is significantly faster than prior cloth simulation systems. We observe almost linear speedups with respect to the number of GPUs.
\end{abstract}




%
%

\begin{teaserfigure}
  \center
   \includegraphics[width=0.95\linewidth]{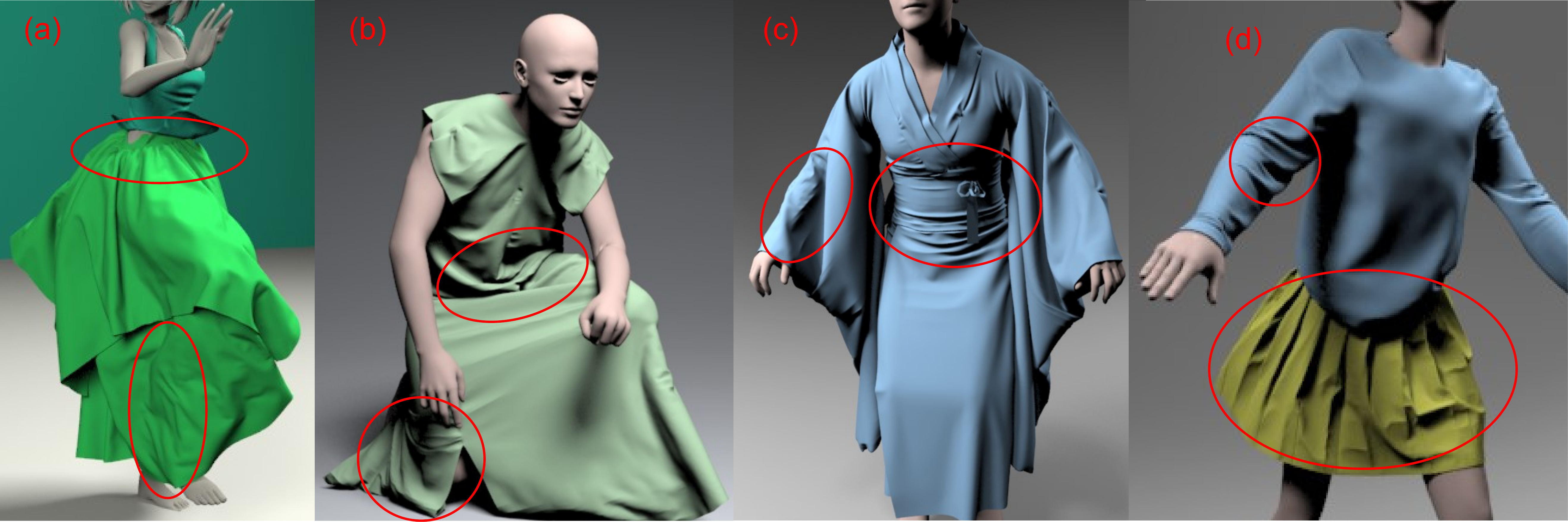}
   \caption{\label{fig:wrinkles}
  {\bf Fast Simulation on Complex Benchmarks:} Our novel multi-GPU based cloth simulation algorithm can  simulate  complex cloth  meshes ((a) Miku with $1.33$M triangles, (b) Kneel with $1.65$M triangles, (c) Kimono with $1$M triangles and (d) Zoey with $569$K triangles) with  irregular shapes and multiple layers at $2-8$ fps on workstations with multiple NVIDIA GPUs. We observe up to $8.23X$ speedups on 8 GPUs. Ours is the first approach that can perform almost interactive complex cloth simulation with wrinkles, friction and folds on commodity workstations. We highlight the areas with detailed wrinkles.}
\end{teaserfigure}

\maketitle


\section{Introduction}
\label{sec:intro}

Cloth simulation is an active area of research in computer graphics, computer-aided design (CAD) and the fashion industry. Over the last few decades many methods have been proposed for solving the underlying dynamical system with robust collision handling. These algorithms are implemented as parts of commercial animation and CAD systems and are widely used for different applications.


The two main components of a cloth simulation system are time integration and collision handling. Each of these steps can be time consuming and many parallel methods have been proposed to exploit multiple cores on a CPU or a GPU for higher performance. In particular, the high number of cores and memory throughput of a GPU have been used to accelerate the overall simulation pipeline~\cite{PKS10,cama16,tang-siga18}. Current single GPU-based cloth simulation systems can achieve $2-8$ fps on small- to medium-resolution cloth meshes with a few hundred thousand triangles. However, their performance is limited by the amount of memory available on a single GPU. Many high fidelity cloth simulation systems use cloth meshes with millions of vertices~\cite{Kutt18,Eberle18} to generate convincing details like wrinkles, friction, folds, etc (See Fig.~\ref{fig:wrinkles} and the video). Furthermore, fast simulation algorithms use different spatial data structures to accelerate time integration or collision detection (e.g., using bounding volume hierarchies or spatial hashing), which considerably increase the memory overhead. It is non-trivial to fit such meshes and the associated data structures on commodity GPUs, which have a few gigabytes memory (e.g., $11$GB memory on a NVIDIA GeForce GTX 1080 Ti GPU).

Many parallel techniques have also been proposed to utilize a large number of CPUs on a cluster~\cite{Selle09,Ni15,Liang18}. However, current CPUs have a lower memory bandwidth and a smaller number of cores than commodity GPUs. Moreover, workstations or computer systems with multiple GPUs are becoming widely available, e.g., NVIDIA DGX/DGX-2 workstations.
As a result, it is useful to design fast parallel multi-GPU algorithms for simulating complex cloth and robustly handling the collisions. In particular, cloth simulation offers many unique challenges in terms of self-collisions or penetration handling that makes it difficult to use commonly used parallel techniques like domain decomposition. One of the major challenges is how to design methods that reduce data transfer between multiple GPUs.

\noindent \textbf{Main Results:}
In this paper, we present a novel multi-GPU based cloth simulation algorithm (P-Cloth) for high resolution meshes. Our approach parallelizes all the stages of the cloth simulation pipeline, including implicit time integration and collision handling. The novel contributions of our work include:
\begin{enumerate}
\item {\bf Pipelined SpMV:} We present a novel sparse matrix-vector multiplication (SpMV) algorithm that can handle dynamic layouts and achieves high throughput  by interleaving the computations and data transfers. We also describe a new work queue generation algorithm for fat-tree interconnect topology that is used to optimize data transfer between different GPUs and improves the overall throughput (Section~\ref{sec:spmv}). We observe $1.4-2.2X$ speedups and improved scalability over prior  multi-GPU SpMV algorithms. 
\item {\bf Dynamic Matrix Assembly:} We present a new technique for matrix assembly and sparse matrix filling that accounts for dynamic contact forces and can be used with a preconditioned conjugate gradient (PCG) solver on a multi-GPU system. The matrix assembly elements are computed in a distributed manner (Section~\ref{sec:matrix}). We observe linear scalability in terms of the number of GPUs. 
\item {\bf Parallel Collision Handling:} We present parallel algorithms for discrete and continuous collision detection using spatial hashing. Our formulation distributes the computations over multiple GPUs, such that the memory overhead of a single GPU is significantly reduced. We propose a parallel non-linear impact zone solver to handle penetrations on multiple GPUs with small data-synchronization overhead (Section~\ref{sec:spatial-hashing}). We demonstrate that our collision detection and response computation scales linearly with the number of GPUs.
\end{enumerate}

These techniques have been integrated and used to perform fast cloth simulation on complex meshes with $0.5-1.65M$ triangles on {two multi-GPU workstations with $4$ NVIDIA Titan Xp GPUs and $8$ NVIDIA Titan V GPUs, respectively.} The memory overhead of P-Cloth for each GPU is about $4-8$ GB and most of the memory are used for the spatial hashing data structure and pairwise collision tests. Prior GPU-based algorithms~\cite{cama16,tang-siga18} are limited to use a single GPU and cannot handle such meshes because the memory overhead can be more than $25$ GB, which exceeds the memory capacity of commodity GPUs. We observe almost interactive performance ($2-5$fps) on these multi-GPU systems.  In contrast, prior distributed methods for cloth simulation that based on matrix-free methods and can take several minutes per frame on meshes with more than $1M$ triangles on multiple Intel Xeon CPUs~\cite{Selle09}. 
Moreover, we observe up to $8.23X$ speedups on 8 NVIDIA Titan V GPUs. We analyze the scalability of each algorithm and observe quasi-linear speedups on overall simulation performance. 
In practice, P-Cloth is the first interactive cloth simulation algorithm that can handle complex cloth meshes on commodity workstations.

%


\section{Prior Work and Background}
\label{sec:related}


\begin{figure}[t]
  	\centering
  	\includegraphics[width=0.95\linewidth]{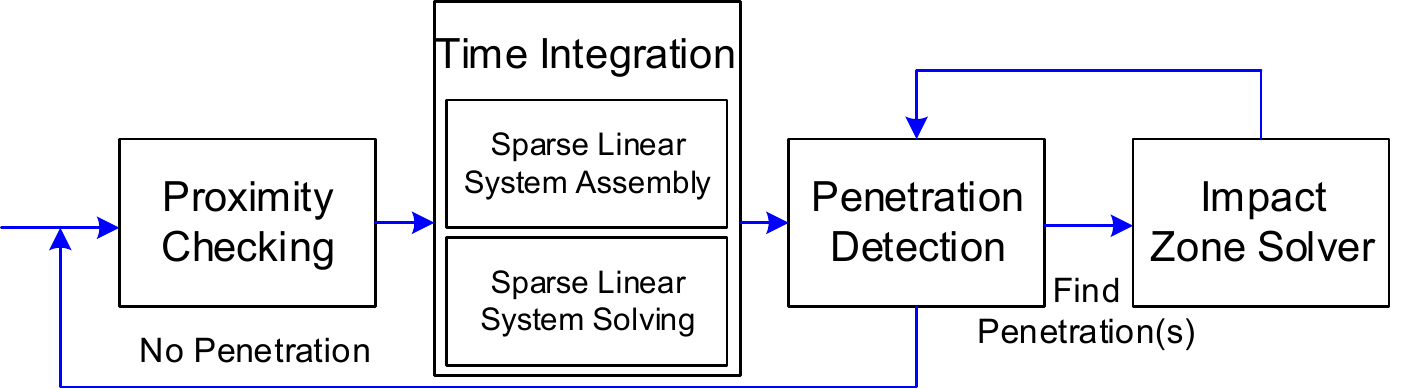}
	\caption{\textbf{Simulation Pipeline:} The pipeline consists of time integration with contact forces, collision detection, and collision response computation. We use well-known implicit time integrator along with contact forces~\cite{Bridson02,Otaduy09}, continuous collision detection for penetration detection~\cite{i3d18}, and impact forces for collision resolution~\cite{Harmon08,tang-siga18}. We present new parallel algorithms that perform these computations on multiple GPUs with reduced data transfers, including dynamic matrix assembly, pipeline implicit integrator, and parallel collision handling.
	}
	
	\label{fig:pipeline}
\end{figure}

\subsection{Time Integrators}
Many integration schemes have been proposed to increase the accuracy and performance of cloth simulation. The earlier schemes were based on explicit Euler integration~\cite{Provot95}, which are quite fast but result in stability problems with large time steps. To overcome stability issues, numerical methods based on implicit Euler integrators~\cite{Baraff1998}, iterative optimization~\cite{Liu13,Liu17}, projective dynamics~\cite{Bouaziz14}, Anderson acceleration~\cite{Peng18}, 
and alternating direction method of multipliers (ADMM)~\cite{Overby17} have been widely used. 
Many local/adaptive techniques have been proposed to  generate the dynamic detail of the cloth and to accelerate the computation~\cite{Yoon10,Narain12,Wang18}. 

\subsection{Collision Detection and Response}
Robust collision handling, including collision detection and response computation, is critical for high-fidelity cloth simulation. Even a single  missed or failed penetration during handling could result in noticeable artifacts~\cite{Bridson02}. There is a considerable literature on accurate collision detection using continuous methods (CCD) and polynomial equation solvers~\cite{Brochu12,Provot97,Tang14,Wang14,MD94,manocha1998solving} to ensure that not a single collision is missed between discrete steps. These algorithms are used for self-collisions and collisions between the cloth mesh and other objects in the scene. Many techniques have been proposed for collision response, including impulse computation~\cite{Bridson02,sifakis08}, constraint solvers~\cite{Otaduy09}, and impact zone methods~\cite{Provot97,Harmon08,tang-siga18}. Other robust methods are based on elastoplastic friction models~\cite{GHFGTT18}, asynchronous contact mechanics~\cite{Vouga2011}, global intersection analysis~\cite{Baraff03,Ye2017,Eberle18}, and volume simulation~\cite{Muller15,Wong18}.

\subsection{Parallel Algorithms}


Given the high complexity of cloth simulation, many parallel techniques have been proposed to accelerate time integration, collision handling, or the entire cloth simulation pipeline. In particular, multiple cores on a single GPU have been used for 
time integration~\cite{cama16,CLMO14} and collision detection~\cite{Govindaraju05,i3d18}. However, these methods assume that the entire cloth mesh and the acceleration data structures  fit into the memory of a single GPU and are limited to small- to medium-sized cloth meshes (e.g. up to a few hundred thousand triangles). Other techniques have been proposed to parallelize the cloth simulation on a CPU cluster~\cite{Selle09,Ni15,Liang18} or a hybrid  combination of CPU and GPU~\cite{PKS10}. Our approach for parallel cloth simulation is designed to exploit the computational capability and high memory bandwidth of GPUs and is complementary to prior parallel simulation algorithms.


Multi-GPU systems have been used for many scientific problems and applications, including  3D finite-difference time domain~\cite{zhou2013multi,shao2017multiple}, stencil computations~\cite{sourouri2015cpu+}, PDE solvers~\cite{malahe2016pde},  fluid simulation~\cite{Hutter14,Liu:2016,chu2017schur}, and material point methods~\cite{WXL2020}. Many of them are based on domain decomposition, which  divides the computational region into sub-regions, then solves each  sub-region independently on 
each GPU. However, these decomposition methods cannot be directly used for robust cloth simulation, where the linear systems for time integration are recomputed during each frame using contact forces~\cite{Otaduy09,cama16}. These forces change dynamically based on the cloth/obstacle mesh configurations and result in new interface regions for the resulting sub-domains.

Sparse matrix-vector multiplication (SpMV) is a time-consuming operator that arises in scientific computation. Many techniques have been proposed to perform parallel SpMV on multiple GPUs for higher throughput~\cite{gao2017novel,guo2016performance}.
However, these methods may not work well in iterative solvers (e.g. conjugate gradient solvers, etc), since the results of each iteration are required to be synchronized among multiple GPUs, i.e. expensive "all-to-all" data transfers.



\subsection{Interconnect between Multiple GPUs}
Current multi-GPU systems or workstations support two kinds of inter-connectivity. The standard systems are based on PCI-e bus, which is a high-speed serial computer expansion bus standard based on point-to-point topology. Some newer systems support NVLink, which is a wire-based communications protocol and is used for data and control code transfer between the GPUs. In practice, NVLink specifies point-to-point connections, which provides $2-3X$ higher bandwidth than PCI-e. 

Current high-end computer workstations equipped with NV-Switches support full NVLink connection with up to $16$ GPUs, allowing all GPUs to communicate with  others without blocking. Other systems typically arrange multiple GPUs in a hierarchical topology, i.e. a binary fat-tree, where the physical distance between a GPU pair can have a noticeable impact on communication efficiency~\cite{faraji2016topology}. Data transfer between a GPU pair with greater physical distance will need to traverse through a higher number of switches and longer paths, and thereby resulting in lower memory bandwidth. Most workstations like NVIDIA DIGITS DevBox use fat-tree topology for multi-GPU interconnect~\cite{NVIDIADevBox}. Our parallel implicit integrator takes GPU topologies and inter-connectivity into account, and strives to cut down data-transfer overhead. 

\subsection{Cloth Simulation Pipeline}

Our basic simulation pipeline is based on implicit time integration, contact force computation, collision detection using spatial hashing, and collision response using a non-linear impact zone solver, as proposed in prior literature~\cite{Baraff1998,Bridson02,Harmon08,Otaduy09,cama16,tang-siga18} (Fig.~\ref{fig:pipeline}).

{We use implicit integrators with contact forces due to its stability, simulation fidelity, and benefits in terms of GPU parallelization~\cite{Baraff1998,Narain12,tang-siga18}. Furthermore, it can be easily integrated with GPU-basd collision handling methods. }
We use a triangle mesh based piece-wise linear elastic model to simulate cloth with non-linear anisotropic deformations. Given a $p$-vertex mesh used to represent the cloth, the overall mesh $Y$ corresponds to a point in a high-dimensional space: $Y=R^{3p}$.
We assume that the initial state of the cloth mesh is penetration-free.

We formulate the dynamical system as the following equation:
\begin{equation}
\mathbf{M}\ddot{\mathbf{u}}=\mathbf{f},
\end{equation}
where $\mathbf{M} \in R^{{3p} \times {3p}}$ is the mass matrices of the vertices, $\mathbf{u} \in R^{3p}$ is the displacement vector of the vertices, $\mathbf{f} \in R^{3p}$ is the force vector computed using the internal and external forces. We perform time integration using the backward Euler method~\cite{Baraff1998}, and approximate the force vector $\mathbf{f}$ at time $t+\Delta{t}$ using a first-order Taylor expansion: 
\begin{equation}
\mathbf{f}(\mathbf{u}_{t+\Delta{t}})=
\mathbf{f}(\mathbf{u}_{t})+\mathbf{J}\cdot(\mathbf{u}_{t+\Delta{t}}-\mathbf{u}_{t}),
\end{equation}
where $\mathbf{J} \in R^{{3p} \times {3p}}$ is the Jacobian matrix of $\mathbf{f}$ evaluated at time $t$.
This results in the following linear equations:
\begin{equation}
\label{eqt:linear-system}
(\mathbf{M}-\Delta{t}^{2}\mathbf{J})\cdot\Delta{\mathbf{v}}=
\Delta{t}\mathbf{f}(\mathbf{u}_{t}+\Delta{t}\mathbf{v}_{t}),
\end{equation}
where $\mathbf{v} \in R^{3p}$  is the velocity vector and $\Delta{\mathbf{v}}$ is the increment of $\mathbf{v}$, which is the unknown variable to be solved.
We solve the linear equations using a preconditioned conjugate gradient (PCG) solver. 
These computations are performed at each time step. The impact zone constraint-enforcement is performed, decoupled from time integration, and is similar to prior collision response algorithms~\cite{Bridson02,Harmon08,tang-siga18}.






\section{Parallel Time Integration on Multiple GPUs}
\label{sec:time-integration}

In this section, we present a novel parallel time integration algorithm that maps well to multiple GPUs. A key computation in the algorithm is to solve the linear system shown in Equation \ref{eqt:linear-system}.
Due to the use of contact forces, the time integration algorithm results in a new linear system during each time step. These contact forces can appear randomly due to the configuration of the cloth or environment objects, which methods based on domain decomposition can not efficiently deal with.
Many conjugate gradient methods have been designed for multi-GPU systems, but are not designed to handle dynamic interface regions~\cite{cevahir2009fast,goddeke2007exploring,muller2014petascale,verschoor2012analysis,kim2011gpu,Marco2010,georgescu2010conjugate}. 
Instead, we present novel multi-GPU parallel algorithms that can solve a sparse linear system with a dynamic layout.
As shown in Fig.~\ref{fig:pipeline}, during each time step we first perform sparse linear system assembly followed by sparse linear system solving.



In this section, we present efficient parallel algorithms for matrix assembly and sparse solvers.
Our formulation treats the overall computational domain as an entirety, and distributes the data and computation tasks among multiple GPUs. In particular, we  parallelize the most computationally expensive operator, sparse matrix-vector multiplication (SpMV) using a novel method called {\em Pipelined SpMV}, as shown in Fig.~\ref{fig:sparse_matrix}(b).
Furthermore, we present a new algorithm for assembling the sparse matrix dynamically during each time step on a multi-GPU system, thereby ensuring that the resulting matrix is compatible with a preconditioned conjugate gradient (PCG) solver. 

\subsection{Pipelined SpMV}
\label{sec:spmv}

We extend {the PCG solver proposed by Tang et al.~\shortcite{cama16}} by performing vector operations and SpMV operations in parallel on multiple GPUs. Vector operations can be easily implemented on multiple GPUs: a vector can be divided into sub-vectors based on the number of GPUs and each GPU handles the corresponding part without any communication with other GPUs. 

There is extensive work on fast implementations of SpMV on GPUs~\cite{filippone2017sparse}, though most of these methods are designed for a single GPU. In the context of multi-GPU systems, one of the main issues is to reduce the communication overhead between GPUs. 
Previous algorithms allocate memory on the host for storing the overall vertor $\mathbf{v}$. During each iteration, the GPUs copy input vectors at the beginning for SpMV operation and send the final results back to the host memory~\cite{cevahir2009fast,verschoor2012analysis,kim2011gpu,yamazaki2015mixed,georgescu2010conjugate}. In practice, waiting for such vector data from other GPUs can be time consuming and it is non-trivial to scale such methods with the number of GPUs. We present a new Pipelined SpMV algorithm that reduces the overhead on multi-GPU systems.

We use the symbol $n$ to denote the number of GPUs in our multi-GPU system.
The matrix $\mathbf{A}$ with dimension $m \times m$ is split into $n$ sub-matrices $(\mathbf{A}_0, \ldots, \mathbf{A}_{n-1})$, each with $m/n$ rows.
Note that vector $\mathbf{v}$ is also divided into $n$ sub-vectors $(\mathbf{v}_0, \ldots, \mathbf{v}_{n-1})$, resulting in the following SpMV computation:
\begin{small}
	\begin{equation}
	\begin{aligned}
	\mathbf{A}\mathbf{v} =
	\left[
	\begin{array}{c}
	\mathbf{A}_{0}\\
	\mathbf{A}_{1}\\
	\vdots\\
	\mathbf{A}_{n-1}\\
	\end{array}
	\right]
	\times
	\left[
	\begin{array}{c}
	\mathbf{v}_{0}\\
	\mathbf{v}_{1}\\
	\vdots\\
	\mathbf{v}_{n-1}\\
	\end{array}
	\right]
	=
	\left[
	\begin{array}{c}
	\mathbf{A}_{0}(\mathbf{v}_{0} + \mathbf{v}_{1} + \ldots + \mathbf{v}_{n-1})\\
	\mathbf{A}_{1}(\mathbf{v}_{0} + \mathbf{v}_{1} + \ldots + \mathbf{v}_{n-1})\\
	\vdots\\
	\mathbf{A}_{n-1}(\mathbf{v}_{0} + \mathbf{v}_{1} + \ldots + \mathbf{v}_{n-1})\\
	\end{array}
	\right].
	\end{aligned}
	\end{equation}
\end{small}
{Moreover, we align $(\mathbf{v}_0, \ldots, \mathbf{v}_{n-1})$ to the same length by filling zeros at the end. }

\begin{figure}[t]
	\centering
	\includegraphics[width=0.85\linewidth]{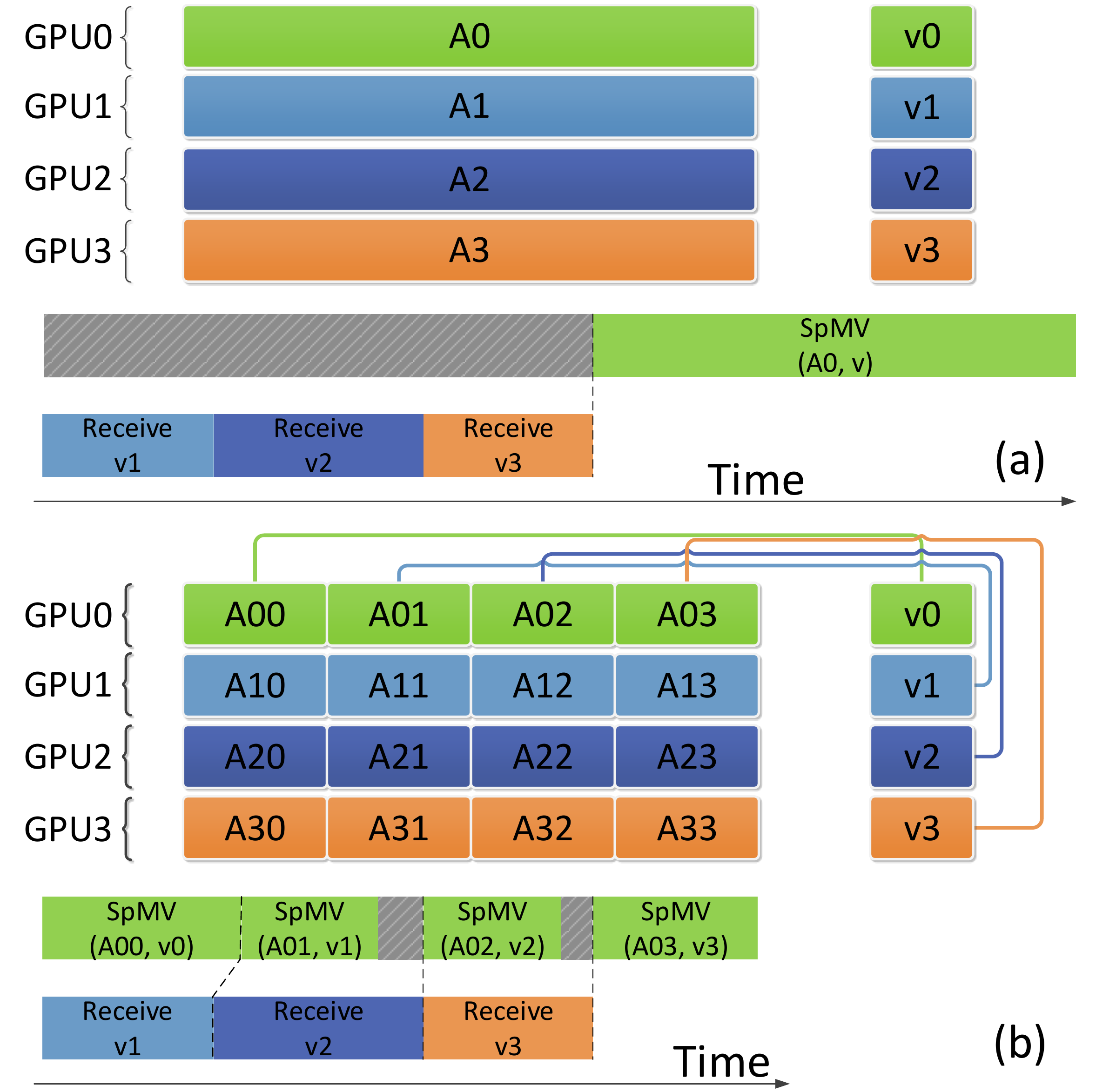}
	\caption{\textbf{Layout of Sparse Matrix on a 4-GPU System}: We demonstrate our approach on a 4-GPU system with each GPU shown using a different color. The bottom timeline shows the computations on $GPU_0$. (a) A straightforward implementation of SpMV~\cite{cevahir2009fast} can lead to substantial waste of computing resources (highlighted in grey color). (b) Our new Pipelined algorithm interleaves the computations and data transfers to achieve better GPU utilization.}
	\label{fig:sparse_matrix}
	\vspace*{-0.12in}
\end{figure}

The SpMV for each sub-matrix can be assigned to a different GPU.
To perform a SpMV computation for each sub-matrix $\mathbf{A}_i$ ($i \in [0,n-1]$), the entire vector $\mathbf{v}$ is required beforehand, causing "all-to-all" data transfer among the GPUs~\cite{Nathan16}. In other words, $GPU_0$ is assigned to multiply $\mathbf{A}_0$ with the entire input vector, only after it receives $\mathbf{v}_1, \ldots, \mathbf{v}_{n-1}$ from $GPU_1,GPU_2,\ldots,GPU_{n-1}$, respectively,  can the SpMV start. The overall SpMV operation can be executed based on the timeline shown in Fig.~\ref{fig:sparse_matrix}(a).


We present a novel pipeline-based parallel algorithm to perform SpMV computation. The key idea is to perform other computations on a GPU, while they are waiting for data from other GPUs corresponding to the last iterative step.
As shown in Fig.~\ref{fig:sparse_matrix}(b), we further split the matrix along its columns, resulting in $n\times n$ sub-matrices $\mathbf{A}_{ij}$, where $i, j \in [0, n-1]$, each with dimension $(m/n) \times (m/n)$, while the SpMV computation is performed in a similar manner. In this case, we obtain:
\begin{small}
	\begin{equation}
	\begin{aligned}
	\mathbf{A}\mathbf{v} &=
	\left[
	\begin{array}{cccc}
	\mathbf{A}_{00}&\mathbf{A}_{01}&\cdots&\mathbf{A}_{0,n-1}\\
	\mathbf{A}_{10}&\mathbf{A}_{11}&\cdots&\mathbf{A}_{1,n-1}\\
	\vdots&&\ddots&\vdots\\
	\mathbf{A}_{n-1,0}&\mathbf{A}_{n-1,1}&\cdots&\mathbf{A}_{n-1,n-1}\\
	\end{array}
	\right]
	\times
	\left[
	\begin{array}{c}
	\mathbf{v}_{0}\\
	\mathbf{v}_{1}\\
	\vdots\\
	\mathbf{v}_{n-1}\\
	\end{array}
	\right]
	\\ &=
	\left[
	\begin{array}{c}
	\mathbf{A}_{00}\mathbf{v}_{0} + \mathbf{A}_{01}\mathbf{v}_{1} + \ldots +\mathbf{A}_{0,n-1}\mathbf{v}_{n-1}\\
	\mathbf{A}_{10}\mathbf{v}_{0} + \mathbf{A}_{11}\mathbf{v}_{1} + \ldots +\mathbf{A}_{1,n-1}\mathbf{v}_{n-1}\\
	\vdots\\
	\mathbf{A}_{n-1,0}\mathbf{v}_{0} + \mathbf{A}_{n-1,1}\mathbf{v}_{1} + \ldots +\mathbf{A}_{n-1,n-1}\mathbf{v}_{n-1}\\
	\end{array}
	\right].
	\end{aligned}
	\label{SpMV}
	\end{equation}
\end{small}

\begin{figure*}[h]
	\centering
	\includegraphics[width=0.98\linewidth]{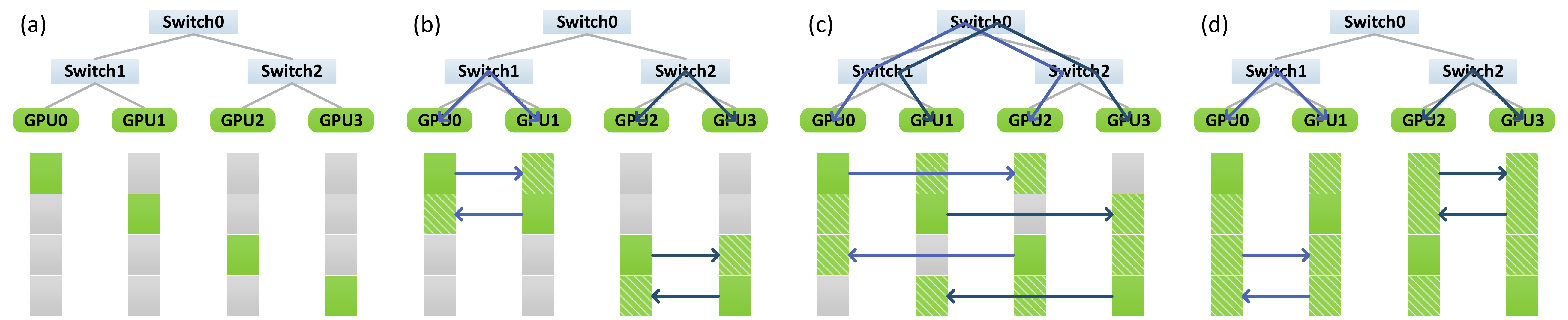}
	\caption{\textbf{Data Transfer Scheme on Fat-tree Topology}: 
		We highlight the performance of our novel work queue generation algorithm on a 4-GPU system with fat-tree topology.
		We highlight the connectivity between different GPUs using various switches and take the interconnectivity into account for data transfer. Fig.(a) shows the initial configuration before the Pipelined SpMV computation, while figures (b), (c), (d) highlight 3 transfer stages, respectively. By performing data transfers one stage at a time, followed by multiplication between the corresponding sub-matrix and sub-vector, our parallel algorithm results in higher throughput of SpMV on a multi-GPU system.
	}
	\label{fig:trans_tree}
\end{figure*}

As the timeline in Fig.~\ref{fig:sparse_matrix}(b) shows, each sub-matrix sub-vector multiplication is performed as soon as the corresponding sub-vector is received from other GPUs. In this formulation, the latency caused by data communication between GPUs is mostly hidden by computation and we obtain better parallel performance for SpMV computation.

\subsubsection{Overhead Analysis}
We analyze the overhead that are associated with our proposed Pipelined SpMV algorithm.
The performance of Pipelined SpMV is determined by the accumulative time costs of computations and data transfers.
For simplicity, the following analysis is based on $GPU_0$ of an n-GPU system.
We use the notation $C_i$ to denote the time cost of computing $SpMV(\mathbf{A}_{0i}, \mathbf{v}_{i})$, $T_i$ for the overhead of receiving sub-vector $\mathbf{v}_{i}$ and $S_i$ for accumulative time by finishing $SpMV(\mathbf{A}_{0i}, \mathbf{v}_{i})$.
For the straightforward implementation (Fig.~\ref{fig:sparse_matrix}(a)), we can easily obtain:
\begin{equation}
	S_i = \sum_{j=1}^{i}{T_j} + \sum_{j=0}^{i}{C_j}.
	\label{overhead-straightforward}
\end{equation}
And for Pipelined SpMV, operator $SpMV(\mathbf{A}_{0i}, \mathbf{v}_{i})$ can be computed only after accomplishing prior computations and receiving necessary data. Therefore, the accumulative overhead is reduced to:
\begin{equation}
	S_i = 
	\begin{cases}
	\max{(S_{i-1}, \sum_{j=1}^{i}{T_j})} + C_i, \quad & (i>0) \\
	Ci.\quad & (i=0)
	\end{cases}
\label{overhead-pipelinedspmv}
\end{equation}

In practice, data communication overhead largely depends on the interconnectivity of the multi-GPU system, which will dominate the $\max{(S_{i-1}, \sum_{j=1}^{i}{T_j})}$ term in Equation~\ref{overhead-pipelinedspmv}.
On advanced workstation which equipped with full NVLink connection such as DGX-2~\cite{NVIDIADGX-2}, data transfer among GPUs is efficient enough to allow completely hiding communication overhead, and therefore, the $\max{(S_{i-1}, \sum_{j=1}^{i}{T_j})}$ term is equal to $S_{i-1}$. In this case, computation resources are fully utilized and our Pipelined SpMV can achieve ideally acceleration, which can be formed as:
\begin{equation}
S_i = \sum_{j=0}^{i}{C_j}. \quad (i>0)
\label{overhead-pipelinedspmv-nvlink}
\end{equation}

On the other hand, on systems using PCI-e bus for inter-GPU data communication, the $\max{(S_{i-1}, \sum_{j=1}^{i}{T_j})}$ term can be approximately considered as $\sum_{j=1}^{i}{T_j}$, due to its limited transfer bandwidth. Then the overhead is:
\begin{equation}
S_i = \sum_{j=1}^{i}{T_j} + C_i. \quad (i>0)
\label{overhead-pipelinedspmv-pcie}
\end{equation}

\subsubsection{Efficient All-to-all Transfer on Fat-tree Topology}

{Recall Equation~\ref{overhead-pipelinedspmv-pcie}, although Pipelined SpMV interleaves computations and communication, in the case that data transfer via PCI-e bus, the latency caused by inter-GPU communication can still be significant and governs the overall performance.}
We present an efficient algorithm to reduce the transfer overhead on multi-GPU systems with fat-tree topology.

Fig.~\ref{fig:trans_tree}(a) shows an example configuration of a fat-tree topology on a 4-GPU system,  where each of the sub-vectors is stored on the according GPU memory. GPUs are interconnected with switches (e.g., PCI-e internal switch, PCI-e host bridge, etc) so that they can communicate with others. GPUs within a shorter distance result in higher memory bandwidth and lower latency~\cite{faraji2016topology}. For example, $GPU_0$ and $GPU_1$ can immediately communicate with each other through $Switch_1$, while communication between $GPU_0$ and $GPU_2$ is more time-consuming than immediate communication (through $Switch_1$, $Switch_0$, and $Switch_2$). We need to take these overhead into account in terms of designing data transfer and work queue generation algorithms.

During Pipelined SpMV computation, each GPU is required to collect sub-vectors from other GPUs.
However, communications between different GPU pairs with overlapped traversal paths can not be performed simultaneously. For example, communication between $GPU_0$ and $GPU_1$ can be blocked while $GPU_0$ is communicating with $GPU_2$, since these two traversal paths share the part between $GPU_0$ and $Switch_0$. 
These constraints can directly impact the performance of data transfers.
Moreover, the execution order of multiplications between sub-matrices and sub-vectors in the SpMV pipeline can be arbitrary. For example, $GPU_0$ can firstly receive $\mathbf{v}_1$ from $GPU_1$ and multiply $\mathbf{A}_{01}$ with $\mathbf{v}_1$. Or $GPU_0$ can receive $\mathbf{v}_3$ and then multiply $\mathbf{A}_{03}$ with $\mathbf{v}_3$.
That means data transfers during Pipelined SpMV can also be reordered for better data traffic.

Another technique used to reduce the overhead of memory access is based on the fact that not all of data transfers via high distance paths are necessary.
During Pipelined SpMV computation, each GPU collects sub-vectors for SpMV, once a GPU obtains data from a remote GPU through a high distance path, this "farther" data can be broadcast to its neighbors.
In order to design an efficient scheme, we take into account all the computations that need to be performed and re-organize the execution order in the overall Pipelined SpMV.

\noindent{\bf Efficient Data Transfer:} Based on these properties, we address the problem of performing $n \times (n-1)$ data transfers on an $n$-GPU system into $n$ work queues, each has length $n-1$. We use the following terminology in the rest of this section.

\begin{enumerate}
	\item A work queue
	$Q_i$
	denotes the execution order of receiving required input vectors on $GPU_i$, each work queue contains $n-1$ queue nodes. Moreover, we extend the notation $Q_i$ to $Q_i(s)$ , where $s$ in the brackets denotes that work queue $Q_i$ is corresponding to a sub-system with the root switch $Switch_s$. 

	\item Each queue node
	$N = (t, v)$
	indicates a data transfer task for the corresponding GPU to $GPU_{t}$, where $v$ represents the sub-vector.
	
\end{enumerate}
We perform data transfers by querying all the work queues in parallel. Transfer tasks sharing the same index in the work queues are included in one \textbf{transfer stage}. During each stage, each of the $n$ GPUs receives a sub-vector from another GPU. We synchronize all the transfer tasks at a given stage and move to the next stage.

We also use the term $Switch$ to represent the corresponding sub-system of specific GPUs in a recursive manner (see Fig.~\ref{fig:trans_tree}).
For $Switch_s$ that interconnects GPUs ranging from $l$ to $r$, its child switches can be considered as two sub-systems, interconnecting GPUs ranging from $l$ to $m-1$ and from $m$ to $r$ respectively, where $m=(l+r)/2$. 

We illustrate our transfer scheme on a 4-GPU system in 
Fig.~\ref{fig:trans_tree}. Fig.~\ref{fig:trans_tree}(a) highlights the initial configuration, and (b),(c),(d) correspond to 3 transfer stages, respectively.
During the second stage shown in Fig.~\ref{fig:trans_tree}(b), $GPU_0$ obtains sub-vector $\mathbf{v}_2$ from $GPU_2$, which can be sent to $GPU_2$ during the third stage, as shown in Fig.~\ref{fig:trans_tree}(c),
so that $GPU_1$ can receive $\mathbf{v}_2$ from $GPU_1$ through lower level $Switch_1$.

Our overall work queue generation algorithm uses a greedy strategy.
In order to compute work queues $Q_l(s), \ldots, Q_r(s)$ of $Switch_s$, we first obtain work queues corresponding to $Switch_{2s-1}$ and \\* $Switch_{2s}$. Next, we interchange data between these two sub-systems via $Switch_s$, ensuring that each sub-vector necessarily traverses through the root switch only once. The rest of the data transfer tasks are delegated to the sub-systems in a recursive manner. This assignment is performed by applying an offset to the vector index $t$ of each queue node $N$ in their work queues.

Our work queue generation algorithm works in a recursive manner using the following three steps:
\begin{enumerate}
	\item Compute work queues $Q_{l}(2s-1), \ldots, Q_{m-1}(2s-1)$ of its left child switch $Switch_{2s-1}$, then append to $Q_l(s), \ldots, Q_{m-1}(s)$.
	
	Then compute work queues $Q_{m}(2s), \ldots, Q_{r}(2s)$ of its right child switch $Switch_{2s}$ and append to $Q_m(s), \ldots, Q_r(s)$.
	\item For each queue $Q_i$ in $Q_l(s), \ldots, Q_{m-1}(s)$, push node $(i, i+m)$ to its back. For each queue $Q_j$ in $Q_m(s), \ldots, Q_r(s)$, push node $(j, j-m)$.
	\item Append work queues of its left child switch $Q_l(2s-1), \ldots, \\Q_{m-1}(2s-1)$ to $Q_l(s), \ldots, Q_{m-1}(s)$, with an offset $m$ added to the sub-vector index of each queue node. Also, append $Q_m(2s), \ldots, Q_r(2s)$ to $Q_m(s), \ldots, Q_r(s)$ with an offset $-m$.
\end{enumerate}

\begin{figure}[h]
	\centering
	\includegraphics[width=1.0\linewidth]{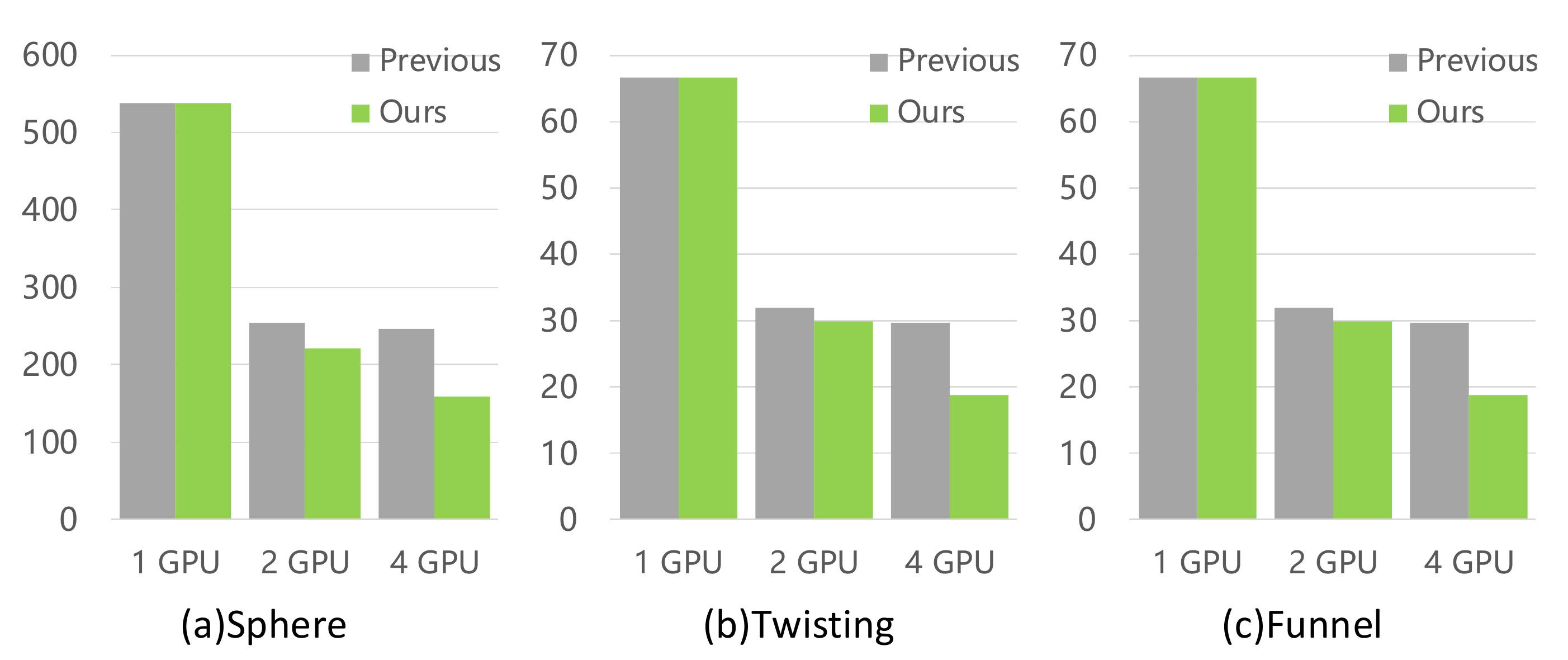}
	\caption{\textbf{Time Cost by SpMV During a Step on Average}: We compare the time cost (ms) using our Pipelined SpMV and a straightforward method on systems with multiple GPUs. Previous approaches~\cite{cevahir2009fast} (in grey) have a higher overhead due to data transfers, while our Pipelined SpMV (in green) offers better performance. As the number of GPUs increase, we observe close to linear speeds using Pipelined SpMV over a  single-GPU implementation~\cite{bell2008efficient}.}
	\label{fig:spmv}
\end{figure}

We assume that the work queues of the root switch are $Q_0(s), \ldots, \\ Q_{n-1}(s)$, where $n$ is the number of GPUs and is a power of 2.
Step (2) optimizes the overall pipeline performance by minimizing the data transfers through higher level switches. 
After step (2), GPUs that lie in the range of the child switches obtain necessary data to deliver the overall vector to each interconnecting GPU. The rest of the transfer jobs are assigned to lower level switches for higher memory bandwidth.
The pseudo-code of our work queue generation algorithm is provided in the supplementary material.
We use these work queues to reduce the data transfer overhead and thereby improve the performance of SpMV based on Equation~\ref{SpMV}.
As shown in Fig.~\ref{fig:spmv}, while the method proposed by Cevahir et al.~\shortcite{cevahir2009fast} achieves $2.2X - 2.3$ speedups a on 4-GPU system, and the performance can hardly further scale from 2-GPU to 4-GPU, our Pipelined SpMV with efficient data transfer scheme scales well among multiple GPUs and further achieves about $2.9X - 3.5X$ speedups .

\subsection{Matrix Assembly on Multiple GPUs}
\label{sec:matrix}

\begin{figure*}[t]
	\centering
	\includegraphics[width=0.9\linewidth]{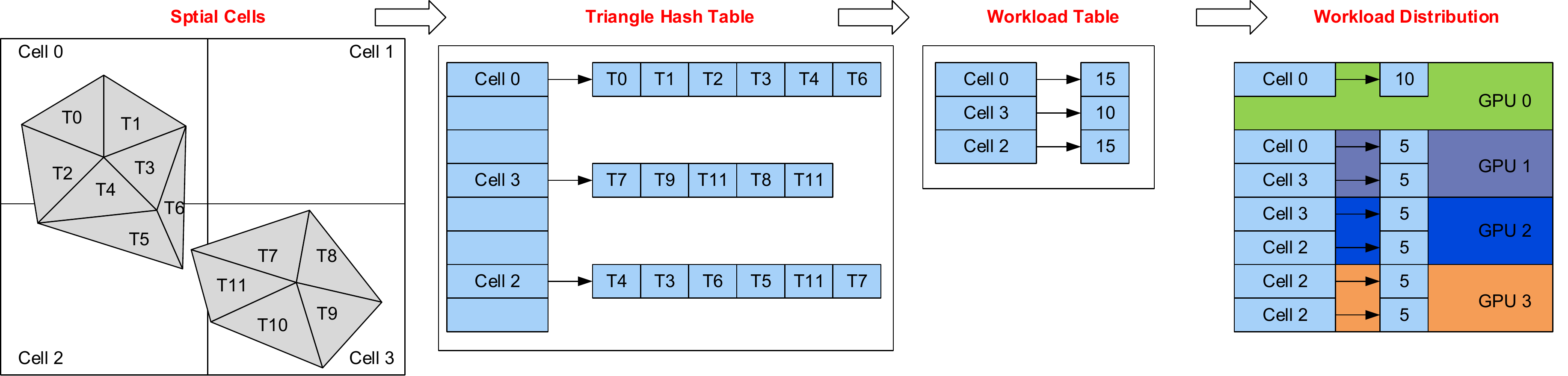}
	\caption{\textbf{Collision Detection with Spatial Hashing:} We extend prior CCD algorithms based on spatial hashing to multi-GPU systems. 
		We construct the triangle hash table and workload table on each GPU, and decompose the tasks based on the workload table. Each GPU performs CCD computation independently using the hash table based algorithm~\cite{i3d18}. Here we highlight the workload distribution on a 4-GPU system, which results lower working set size and almost linear speedups. }
	\label{fig:sh-pipeline}
\end{figure*}

The sparse linear matrix assembly is performed based on proximity computations and contact forces, as shown in Fig.~\ref{fig:pipeline}. Our goal is to design an efficient matrix assembly algorithm for multiple GPUs, that is compatible with the SpMV algorithm described above. 
As discussed above, we split the sparse matrix into $n \times n$ blocks, $n$ row blocks for each of the $n$ GPUs.
Our goal is to isolate the matrix-vector multiplication computation to avoid bottlenecks during inter-GPU communication.

In order to perform matrix assembly, we mark the external and internal forces~\cite{Bridson02,Otaduy09} that are acting on the cloth as the assembly elements, which will form $\mathbf{f}$ and $\mathbf{J}$ in Equation~\ref{eqt:linear-system}. We first allocate memory on each GPU and then distribute
all the assembly elements according to the GPU that they belong to.  
For each GPU, we calculate its assigned assembly elements and fill in the resulting values in the corresponding matrix blocks.
We perform this computation in parallel on multiple GPUs.

\subsubsection{Assembly Elements Distribution}

{We identify the GPUs that the vertices belong to and distribute the elements accordingly.}
Then each GPU fills the matrix based on its memory allocation, forming a slice of the overall $n \times n$ blocks. However, this approach requires data synchronization between the GPUs and rearranges the matrix for compatibility with the PCG solver. 
Furthermore, GPUs need to allocate memory buffers for the incoming data, which has additional overhead.

In order to perform these computations in an efficient manner, we preprocess all the assembly elements by grouping them according to which vertices are needed to perform the relevant computation. In other words, GPUs collect the required assembly elements beforehand according to their entries in $\mathbf{f}$ and $\mathbf{J}$ in Equation~\ref{eqt:linear-system}, then perform the relevant calculations before filling values into the $n$ row blocks.
Finally, the resulting $n\times n$ matrices $\mathbf{A}_{ij}$ ($i, j \in [0, n-1]$) on each of the $n$ GPUs, are used as immediate input by the PCG solver.

\subsubsection{Sparse Matrix Filling}

The matrix assembly algorithm proposed by Tang et al.~\shortcite{cama16} is designed for a single GPU-based simulation and stores the sparse matrix with a block compressed sparse row (BSR) format.
Given the fact that SpMV is the only operator we perform on the sparse matrix, we use the ELLPACK/ITPACK (ELL) format~\cite{grimes1979itpack}, as it offers better performance on SpMV~\cite{bell2008efficient}.
Moreover, we extend ELL to block ELL (BELL) to reduce the overhead of memory access, with a  $3 \times 3$ block size in a structure of array (SOA) form.
The matrix values corresponding to each row are stored in the \textit{Value Table}, while the \textit{Index Table} stores the corresponding column index of non-zero values.
After computing the distributed assembly elements, each GPU can allocate memory and runs the filling algorithm for each matrix block in parallel, using the following steps: 
\begin{enumerate}
	\item {\textbf {Index Table Allocating:}} We count the required memory space for each GPU during the distributing stage. We allocate memory for the \textit{Index Table} according to that.
	\item {\textbf {Index Filling:}} To fill \textit{Index Table}, we scan the distributed assembly elements and fill the column indices in the corresponding rows, using atomic operators to avoid conflicts.
	\item {\textbf {Index Compacting:}} Note that there can be duplicated indices, because multiple forces can be acting on one vertex. We sort each row in the \textit{Index Table} and remove duplicated indices, one GPU thread for each row.
	\item {\textbf {Value Table Allocating:}} We allocate memory for the \textit{Value Table} according to the maximum of the compacted row length.
	\item {\textbf {Value Filling:}} We calculate all distributed assembly elements, and fill the results in the \textit{Value Table} using atomic operators. The value entries in the table are based on the compacted \textit{Index Table} because their memory layouts are identical.
\end{enumerate}

Since our matrix assembly algorithm is data-independent, each GPU performs these computations locally without any inter-GPU communication.
Therefore the computational resources of multiple GPUs are well utilized.
The pseudo-code of our sparse matrix filling algorithm is described in  the supplementary material.

\begin{figure}[h]
	\centering
	\includegraphics[width=0.9\linewidth]{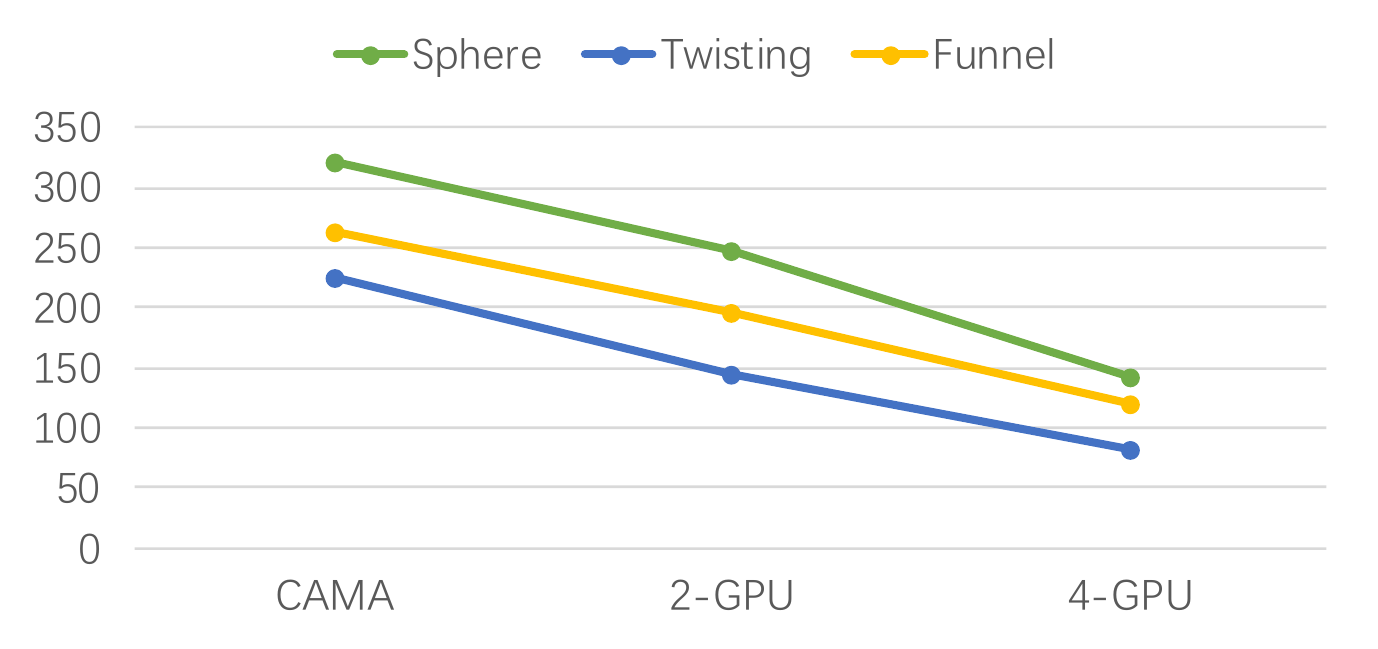}
	\caption{\textbf{Performance comparison with CAMA}: We highlight the time cost (ms) of single-GPU implementation~\cite{cama16} and our parallel martrix assembly algorithm with 2 GPUs and 4 GPUs, respectively. In practice, our matrix assembly algorithm scales well across multiple GPUs.}
	\label{fig:matrix_assembly}
\end{figure}

As shown in Fig.~\ref{fig:matrix_assembly}, our matrix assembly algorithm scales well across multiple GPUs and achieves $2.2X - 2.7X$ speedups on 4-GPU system compared with CAMA~\cite{cama16}.


\section{Parallel Collision Handling: Multiple GPUs}
\label{sec:spatial-hashing}
Collision handling is regarded as one of the major efficiency bottlenecks in cloth simulation, especially for high-resolution cloth meshes.
Prior methods can take up to $82\%$ of the total running  time and have a high memory overhead
, e.g.,  $11$ gigabytes~\cite{i3d18}, to store collision handling related data structures. In this section, we present a parallel algorithm for efficient collision handling on multiple GPUs such that the memory overhead on each GPU is significantly reduced.

\subsection{Collision Handling Pipeline}
Most prior GPU methods for collision detection are based on bounding volume hierarchies (BVHs)~\cite{cama16} or spatial hashing~\cite{PKS10}. Spatial hashing based methods~\cite{tang-siga18,i3d18} offer improved performance in terms of GPU parallelization due to  smaller memory overhead and simpler computation.
Our multi-GPU based collision handling algorithm extends the pipeline of Tang et al.\shortcite{tang-siga18,i3d18} by efficiently distributing the collision detection and response computations on multiple GPUs. One of our goals is to ensure load balancing between the GPUs and minimize the data synchronizing/transferring costs among all the GPUs. Our parallel scheme uses a novel workload scheme that results in more equal distribution. This reduces data synchronization and results in improved performance.

\subsection{Parallel Collision Detection}

For collision detection (both discrete and continuous computation), we first construct the spatial hashing table on-the-fly followed by task decomposition. 
We use the workload table to count computing loads, then distribute all the computing loads evenly on different GPUs. Computing related to one cell can be allocated to multiple GPUs. The algorithm is extended from prior single GPU based CD algorithm~\cite{tang-siga18,i3d18}, and differs from conventional multi-GPU based parallel CD algorithms which distribute workloads based on the cells. Our algorithm has the benefit that can distribute the workloads equally even for the configurations {(e.g. Benchmark Sphere and Benchmark Funnel in the video)} that 
many triangles converge to 
some spatial areas (which makes some cells have much more computing load than others). These configurations are hard to parallelize with conventional methods. After the task decomposition, each GPU performs the collision tests in parallel. The resulting pipeline is shown in Fig.~\ref{fig:sh-pipeline}.

\subsection{Parallel Penetration Handling}

\begin{figure}[t]
  	\centering
  	\includegraphics[width=1\linewidth]{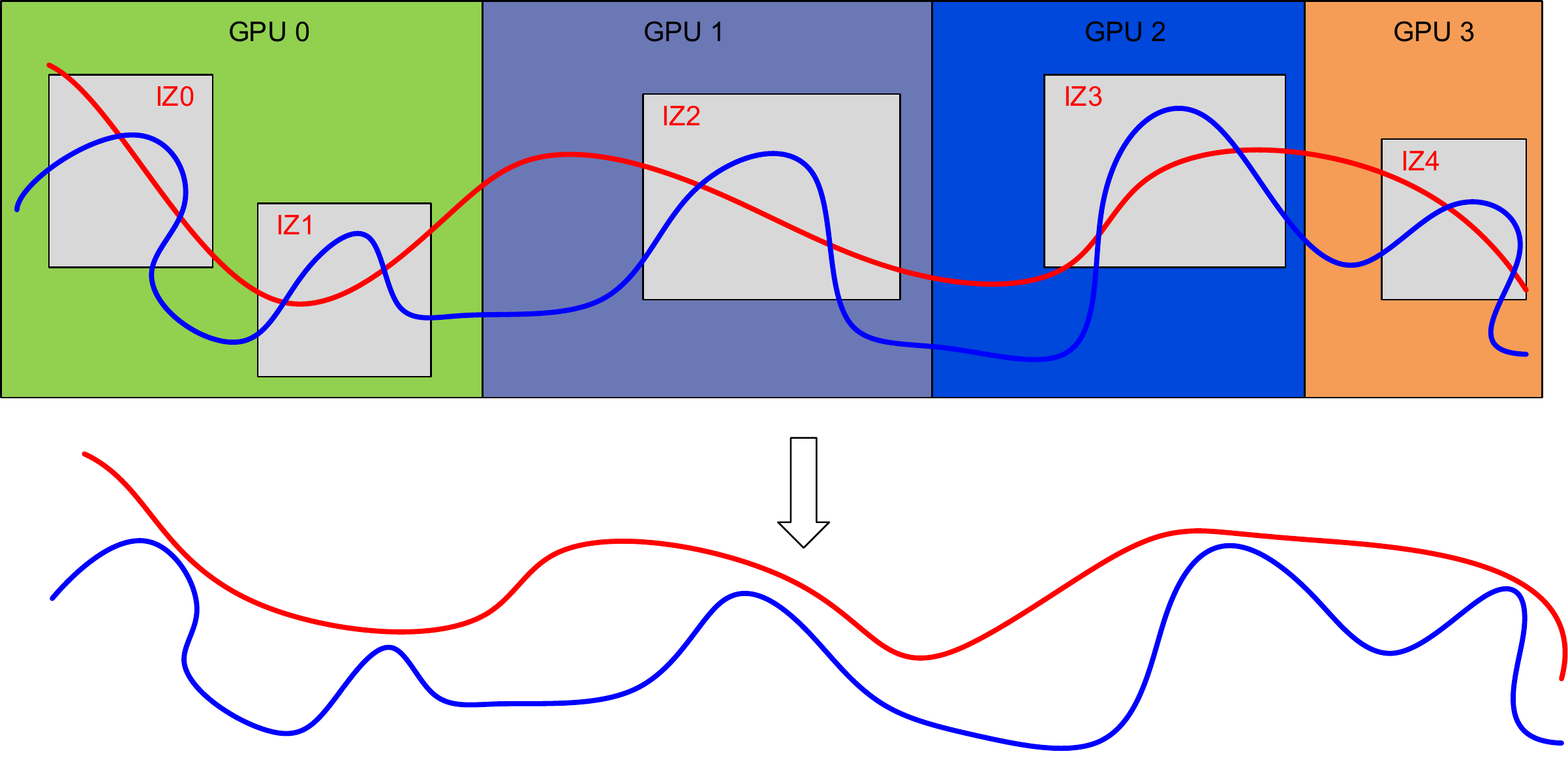}
	\caption{\textbf{Penetration Solving with Multiple GPUs:} We highlight the approach for two colliding meshes, shown in red and blue, respectively.  Multiple penetrations are computed with parallel CCD algorithm (top). All these penetrations are classified as impact zones ($IZ0, \ldots, IZ4$) and  are resolved on multiple GPUs in parallel using a non-linear impact zone solver.}
	\label{fig:penetration-solver}
\end{figure}

 A master GPU is used as an impact zone collector and used for task distribution among all the GPUs.  A GPU thread block is used to solve one impact zone, as shown in Fig.~\ref{fig:penetration-solver}.
 The impact zones are distributed among all the GPUs, and each GPU uses many threads to solve the assigned impact zones.
 For any colliding meshes, including self-collisions, multiple penetrations are computed using CCD. Next, these penetrations are classified as impact zones and solved on multiple GPUs in parallel. A final penetration-free status is computed in an iterative manner for each impact zone.
 
 {We incorporate proximity forces into implicit time integration process. This is different from~\cite{Bridson02}, which is a decoupled solution in which proximities and penetrations are handled separately from time integration. Our integrated approach results in much less penetrations than the decoupled solution, and provides improved stability and fidelity, similar to ~\cite{cama16,tang-siga18,i3d18}.}

 {We only use SpMV  for time integration. The remaining penetrations are handled using inelastic impact zones~\cite{Harmon08}, with a non-linear impact zone solver (with augmented Lagrangian method). Our impact zone solver is implemented as a GPU-optimized gradient descending method~\cite{tang-siga18}. We mostly observe few or narrow penetrations, so the impact zone solver converges in a few iterations ($\leq 5$). We perform a global synchronization among all GPUs after resolving all penetrations. Given few/narrow penetrations, this involves transferring a small amount data ($< 10K$) of adjusted vertices, so the data transfer cost is relatively small.} 

\subsection {Load Balancing}
As shown in Fig.~\ref{fig:sh-pipeline}, the on-the-fly construction {of spatial hashing related data structures} starts with a {\em Triangle Hash Table} based on the distribution of triangles among all spatial cells. 
This process is done on-the-fly for every frame. So it can be used for cloth meshes with dynamic topologies. Next, we  perform collision tests, count how many triangle-triangle tests need to perform for each cell, and generate a {\em Workload Table} based on the count. Finally, we perform {\em load distribution} 
and ensure that the GPU has almost the same number of triangle-triangle test tasks. We avoid any data transfers between the GPUs by performing the construction computation on all the GPUs. As a result, each GPU maintains a copy of its own {\em Triangle Hash Table} and {\em Workload Table}. After the construction process, the collision queries are performed in parallel on all the GPUs.

\begin{figure}[]
  	\centering
  	\includegraphics[width=1\linewidth]{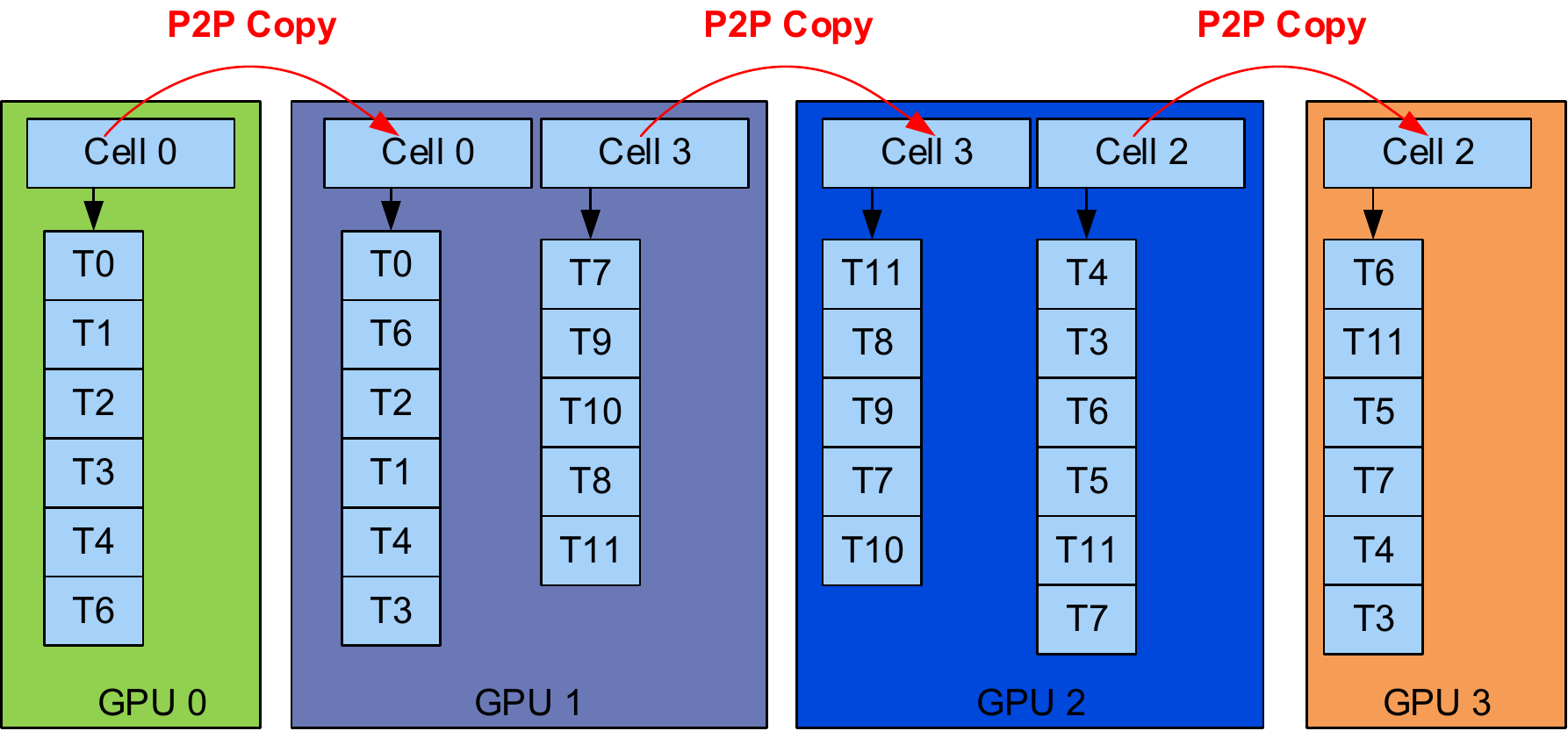}
	\caption{\textbf{Spatial Hashing Data Synchronization among Different GPUs:} Only those cells shared by two GPUs need to be synchronized with P2P data copy on fat-tree topologies.}
	\label{fig:sh-sync}
\end{figure}

\noindent {\bf Data Synchronization:}
Most of the collision detection computations are performed independently on different GPUs. The 
need for data synchronization arises for cells that are shared by two or more GPUs and have the same triangle primitives. For example, the cell $Cell_0$ is used both by $GPU_0$ and $GPU_1$, and its data is stored on both the GPUs (see Fig.~\ref{fig:sh-sync}). In total, only a maximum of $n-1$ GPUs need to be synchronized. 
Overall, such data synchronization has minimal impact on the overall performance since the data transfer is only need for those cells shared by GPUs. 

\begin{figure}[t]
  	\centering
  	\includegraphics[width=0.9\linewidth]{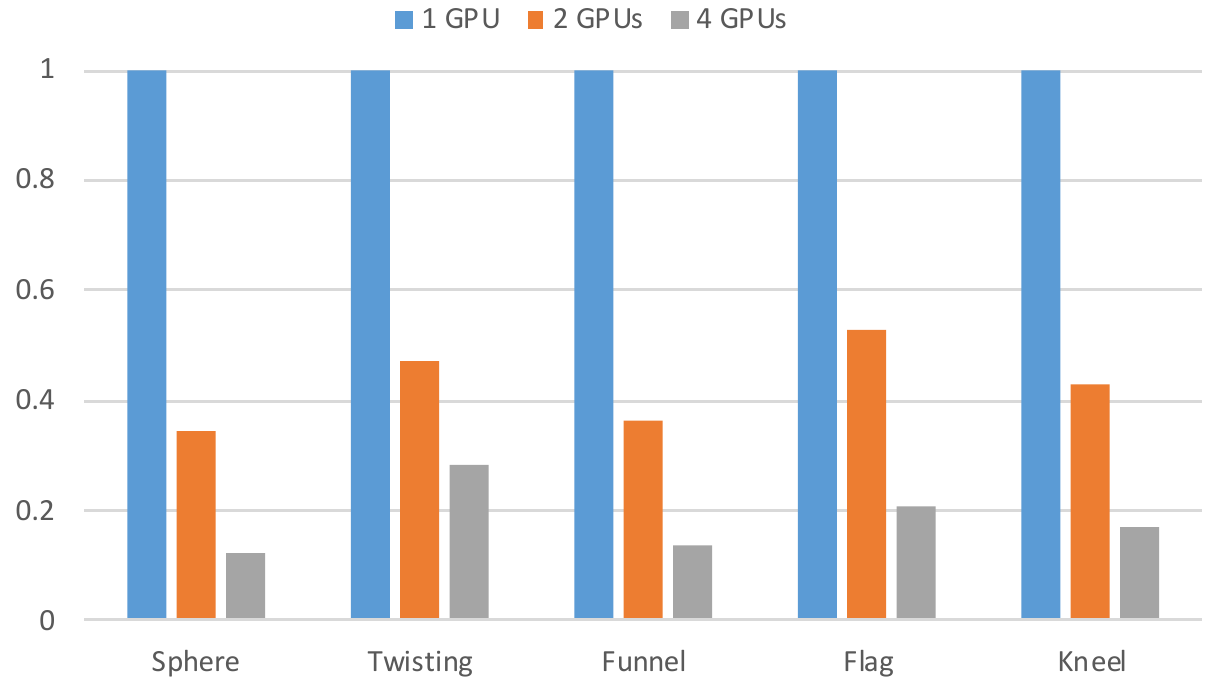}
	\caption{\textbf{Running Time of Parallel Collision Detection with Different Number of GPUs:}  The single GPU configuration is based on I-Cloth algorithm~\cite{tang-siga18} and the 2-GPU and 4-GPU configurations are correspond to our parallel collision detection  algorithm. We observe almost linear speedup as the number of GPUs increase for these benchmarks: Sphere ($700$K triangles), Twisting ($550$K triangles), Funnel ($550$K triangles), Flag ($1.2$M triangles), and Kneel ($1.65$M triangles).}
	\label{fig:collision-time-1-2-4}
\end{figure}

\noindent {\bf Linear Acceleration:} As shown in Fig.~\ref{fig:collision-time-1-2-4}, our load balancing and data synchronization methods result in an almost linear speedup with the number of GPUs. We highlight the results for several complex benchmarks: Sphere ($700$K triangles), Twisting ($550$K triangles), Funnel ($550$K triangles), Flag ($1.2$M triangles), and Kneel ($1.65$M triangles).




\section{Implementation and Results}
\label{sec:results}


We have implemented our parallel cloth simulation algorithm (P-Cloth) and evaluated its performance on {two multi-GPU workstations}. Each of them has 2 Intel Xeon E5-2643 v4 CPUs with 3.40 GHz base frequency, 64 GB system memory. One of them consists of $4$ NVIDIA Titan Xp GPUs (3840 CUDA cores and 12 GB memory per GPU) and the other has $8$ NVIDIA Titan V GPUs (5120 CUDA cores and 12 GB memory per GPU). We run several complex benchmarks on these workstations by varying the number of GPUs to test the parallel performance of P-Cloth.
Our implementation is based on CUDA toolkit 10.0/gcc/Ubuntu 16.04 LTS as the underlying development environment. We use single-precision floating-point arithmetic for all the computations on GPUs.
{Most GPU operations of our implementation are performed in an asynchronous manner for better resource utilization. We use {\em cudaMemcpyPeerAsync} for inter-GPU communications, {\em stream} to overlap computation and data transfers, and {\em cudaStreamSynchronize} for synchronization. The data transfer schedule is controlled by CPU, based on the work queues (Section~\ref{sec:spmv}). }

\begin{figure*}[t]
\centering
\includegraphics[width=0.9\linewidth]{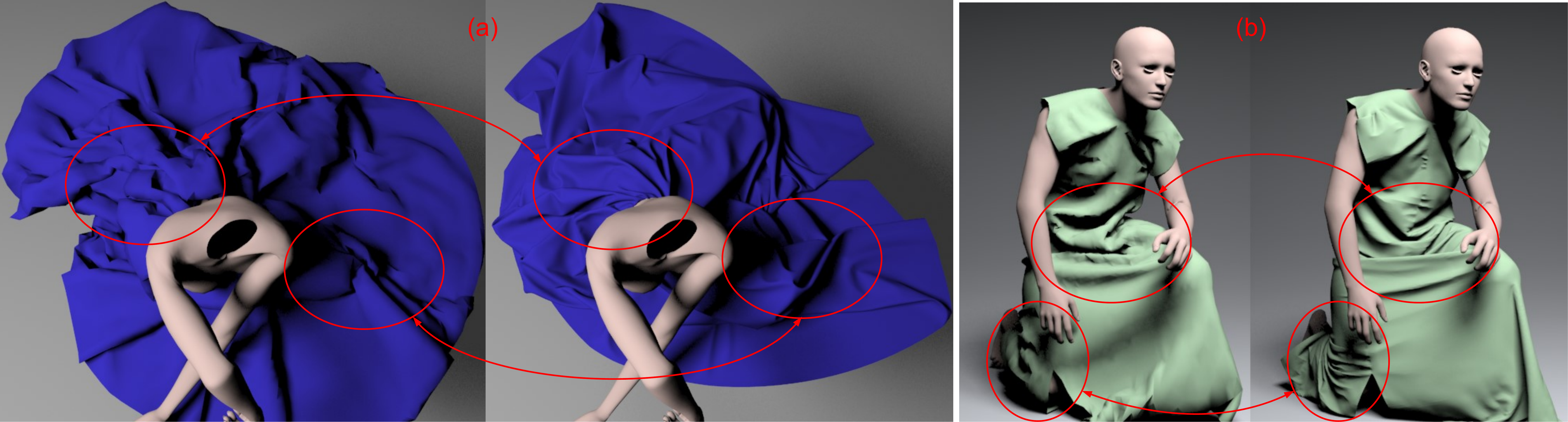}
\caption{ {\bf Resolution Comparison:}  We use different resolutions for the two benchmarks, and highlight the detailed wrinkles and folds for the complex cloth with the higher resolution. The Princess (a) is with $10K$ and $510K$ triangles, and the Kneel (b) is with $3.1K$ and $1.65M$ triangles.}
\label{fig:resolution-compare}
\end{figure*}

We use various benchmarks for regular-/irregular-shaped cloth simulation: 
\begin{itemize}
    \item {\bf Miku:} A dancing girl wearing a ruffled, layered skirt ($1.33$M triangles, Fig.~\ref{fig:wrinkles}(a)).
    \item {\bf Kneel:} A knight is slowly kneeling down  ($1.66$M triangles, Fig.~\ref{fig:wrinkles}(b)).
    \item {\bf Kimono:} A lady bowing with beautiful kimono ($1$M triangles, Fig.~\ref{fig:wrinkles}(c)).    
    \item {\bf Zoey:} A hip hop dancer wearing a pullover and a short skirt with many pleats, ($569K$ triangles, Fig.~\ref{fig:wrinkles}(d)).
    \item {\bf Princess:} A dancer sits on the ground and generates complex folds and wrinkles on the dress ($510$K triangles, Fig.~\ref{fig:resolution-compare}(a), right).
    \item {\bf Andy:} A boy wearing three pieces of clothing (with $538K$ triangles) is practicing Kung-Fu (video). 
    \item {\bf Flag:} A flag with $1.2$M triangles is waving in the blowing wind (video). 
    \item {\bf Sphere:} Three pieces of hanging cloth with $700K$ triangles are pushing by a forward/backward moving sphere (video).
    \item {\bf Funnel:} Three pieces of cloth with  $550K$ triangles is falling into a funnel and folding to fit into it (video).
    \item {\bf Twisting:} Three pieces of cloth with $550K$ triangles that twist considerably, as the underlying ball rotates (video).
    \item {\bf Sphere-1M:} One piece of hanging cloth with $1$M triangles is pushing by a forward/backward moving sphere  (video).
\end{itemize}
These are complex benchmarks with multiple pieces, layers, and wrinkles, which result in a high number of collisions. Prior methods based on a single GPU do not have sufficient memory to store these meshes and the acceleration data structures.  Our P-Cloth algorithm can handle inter- and intra-object collisions reliably and efficiently on the complex benchmarks (see video). The overall accuracy of P-Cloth is the same as that of I-Cloth~\cite{tang-siga18}, as we use the similar geometric and numeric algorithms to compute the contact forces, implicit solvers and non-linear impact zone solver. {Our simulator can support cloth with different physical parameters corresponding to various stretching/bending deformations under external forces~\cite{WHM11}. In our benchmarks, we assign the cloth's material property as the mixture of $60\%$ cotton and $40\%$ polyester.}

\begin{table}[t]
	\centering
	\includegraphics[width=0.85\linewidth]{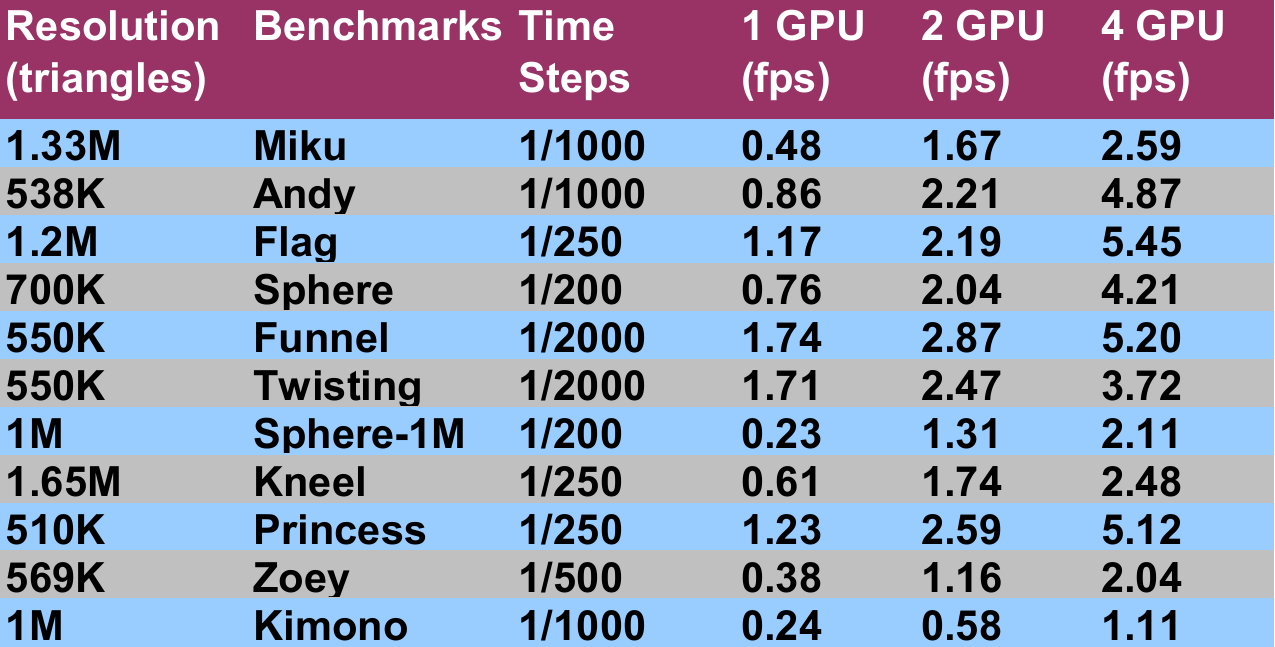}
	\caption{{\bf Performance Comparison:} 
	This table highlights the performance of  P-Cloth for various benchmarks by varying the number of GPUs in the multi-GPU system. We report average FPS on various benchmarks. The speedups over the single-GPU implementation are up to $2.7X$ and $5.6X$, with 2 GPUs and 4 GPUs, respectively. 
	}
	\label{tbl:performance}
\end{table}



\begin{figure}[t]
\centering
\includegraphics[width=0.95\linewidth]{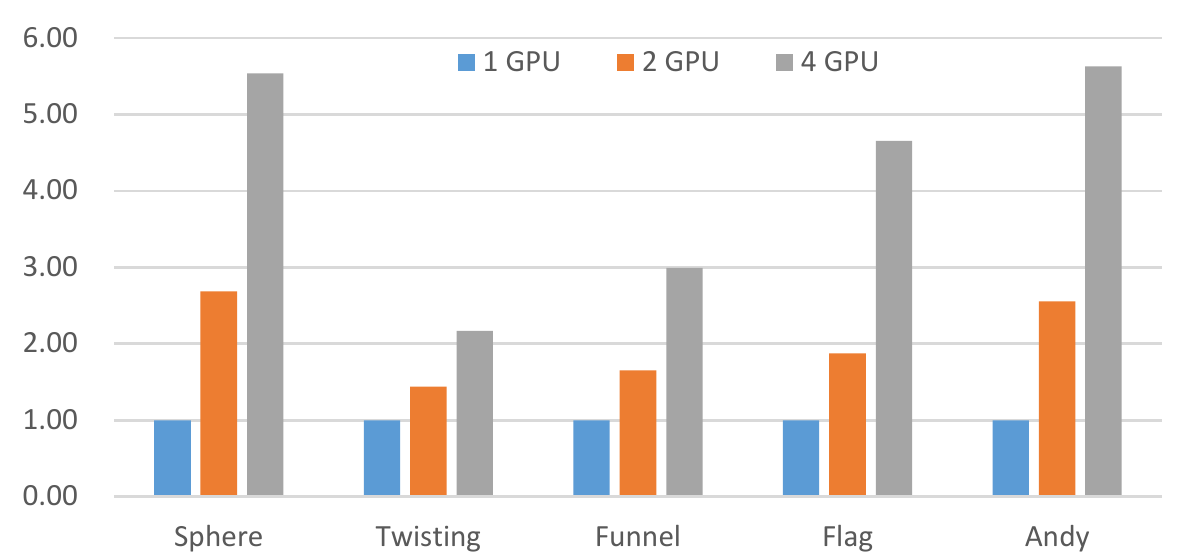}
\caption{ {\bf Parallelization Speedups:} We highlight the speedups using  2 or 4 GPUs for different benchmarks using our P-Cloth algorithm. We observe almost linear speedups, i.e., up to $1.4 - 2.6X$ on 2 GPUs and $2.1 - 5.6X$ on 4 GPUs. The superlinear speedups are due to better memory performance and the reduced working set size.}
\label{fig:speedups}
\vspace{-0.6cm}
\end{figure}

Table~\ref{tbl:performance} shows the mesh resolutions and time step sizes used for different benchmarks. We also highlight the performance of P-Cloth on these benchmarks. This includes the average FPS of P-Cloth with different numbers of GPUs. These results demonstrate that P-Cloth scales well on multiple GPUs and that the performance is almost a linear function of the number of GPUs. Compared with the performance on a single GPU, we observe significant speedups, i.e., up to $2.7X$ on 2 GPUs and up to 
$5.6X$ on 4 GPUs (Fig.~\ref{fig:speedups}). The $5.6X$ speedup is  obtained on the Flag benchmark with very high mesh resolution. This super-linear speedup is due to better memory bandwidth performance and the cache utilization. Since P-Cloth distributes the model and acceleration data structures over different GPUs, the memory overhead and the working set on each GPU for P-Cloth is much lower than a single GPU-based algorithm like I-Cloth~\cite{tang-siga18}. This results in higher memory bandwidth performance and more cache hits for P-Cloth. For benchmark Sphere-1M, we obtained $5.7X$ and $9.2X$ speedups with 2 GPUs and 4 GPUs, respectively. These speedups are achieved due to  the limited GPU memory size on the single GPU. The single-GPU implementation has to split the full working set into several smaller batches, and processes them separately. On the other hand, when we use 2/4 GPUs, each GPU has a much smaller working set, and can process them in parallel. As a result, we observe super-linear speedups for these benchmarks.

\begin{figure}[t]
\centering
\includegraphics[width=0.95\linewidth]{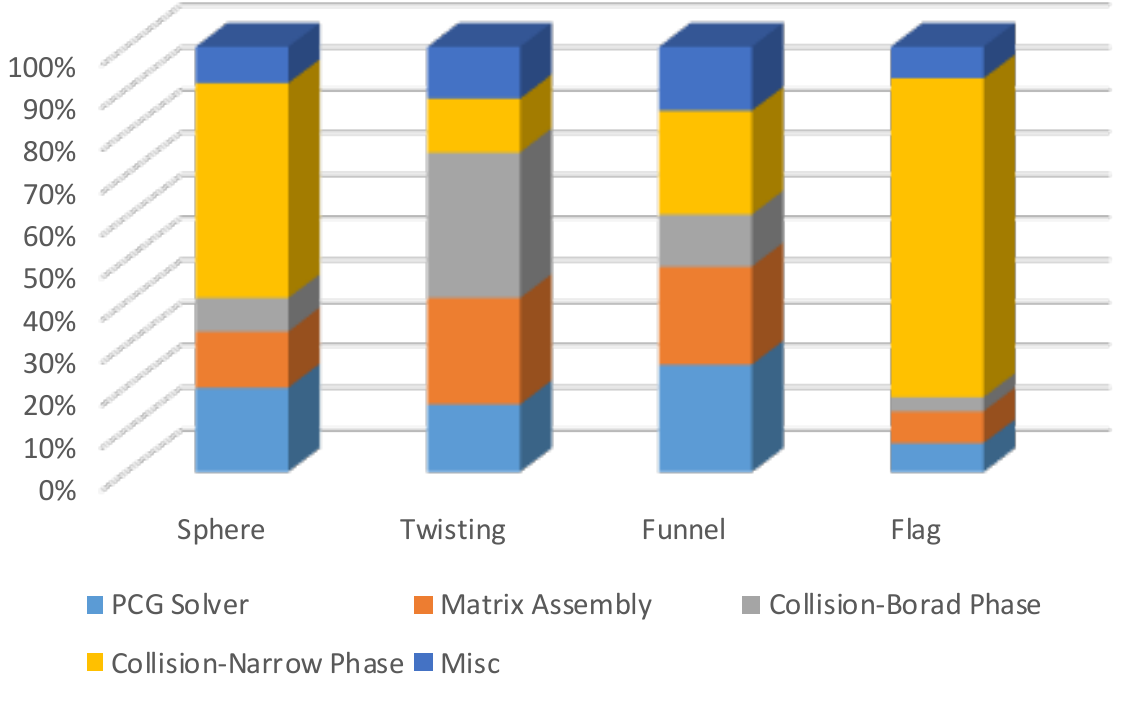}
\caption{ {\bf Run-time Performance Analysis:} We highlight the running time ratios of different computing stages of the cloth simulation pipeline: time integration, broad phrase testing, narrow phrase testing, and penetration handling. The performance data are collected by running P-Cloth on all the benchmarks on the 4-GPU workstation. As shown in the figure, time integration takes almost constant running time for all time steps. In practice, collision detection (including broad phrase and narrow phrase) is the major bottleneck, especially when the cloth is tangled and results in a high number of potential self-collisions.}
\label{fig:runtime-ratios}
\end{figure}

\noindent {\bf Running Time Ratios:}
Figure~\ref{fig:runtime-ratios} shows the running time ratios of different computing stages: time integration, broad phrase testing, narrow phrase testing, and penetration handling. These data are collected by running P-Cloth for all the benchmarks. As shown in the figure, time integration takes almost constant running time for all the time steps. Collision detection (broad phrase and narrow phrase) and penetration handling appear to be the most computationally expensive parts, especially when the cloths are tangled.

\begin{figure}[b]
\centering
\includegraphics[width=1\linewidth]{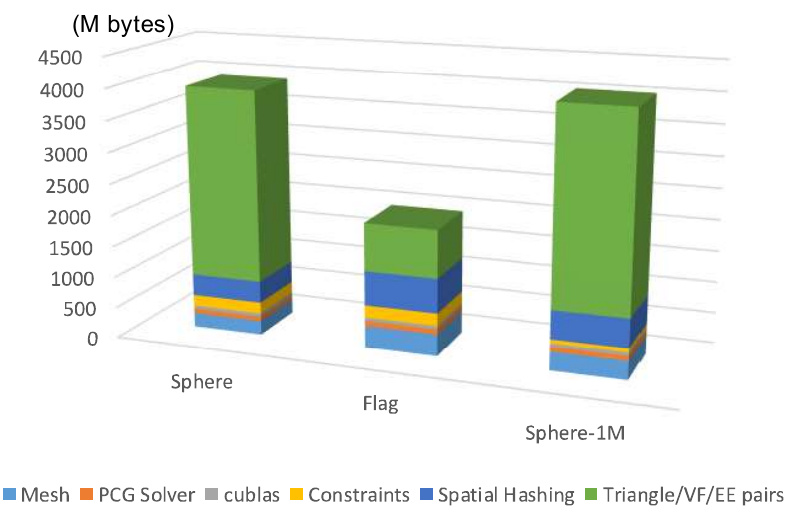}
\caption{ {\bf Memory Overhead:} We highlight  memory overhead of P-Cloth on the  4-GPU workstation. This figure highlights the average memory overhead on each GPU for three complex benchmarks: Sphere, Flag, and Sphere-1M. Most of the memory overhead corresponds to the spatial hashing data structure  and the pairwise triangle/VF/EE CCD tests. It is not possible to run this simulation on a single Titan XP GPU without splitting into sub-batches, as the collision detection algorithm needs more than $12$ GB memory. }
\label{fig:memory-ratio}
\end{figure}

\begin{figure}[b]
\centering
\includegraphics[width=0.9\linewidth]{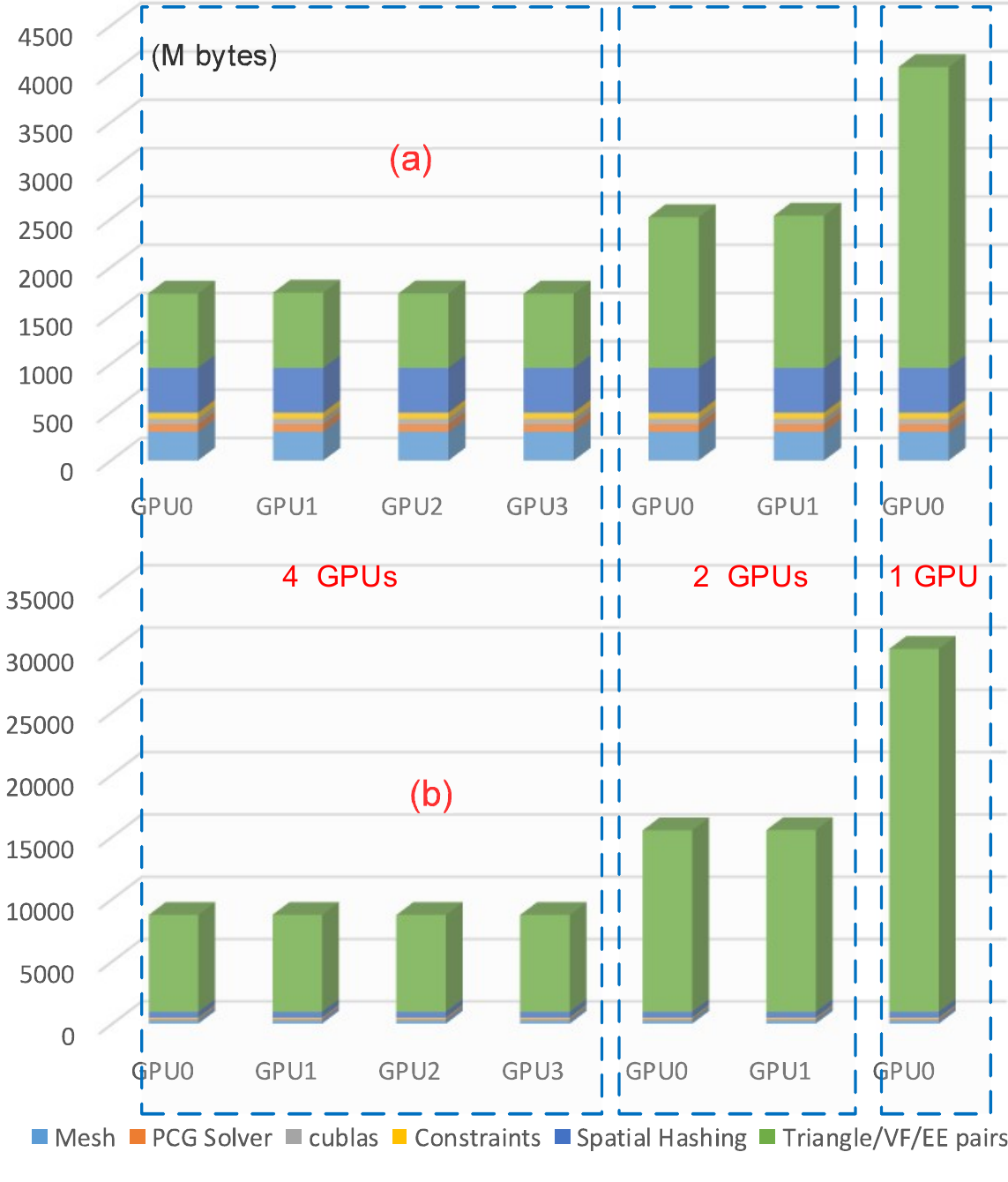}
\caption{ {\bf Memory Overhead vs. Number of GPUs:} We highlight  the memory overhead (per GPU) of P-Cloth versus the number of GPUs used for simulation for Sphere-1M benchmark: (a) Highlights the average memory utilization by different data structures for a frame without many collisions; (b) Highlights the memory overhead for a frame with a high number of potential self-collisions. This results in a very high number of pairwise tests after spatial hashing. For this benchmark, a single-GPU based algorithm would require about $29$ GB and 2-GPU implementation of P-Cloth would need $15$ GB on each GPU.  On a 4-GPU system, P-Cloth takes up to $8$ GB for these challenging close proximity configurations and able to perform the simulation at almost interactive rates. The reduced memory  overhead on a multi-GPU system significantly increases the frame rate and performance.}
\label{fig:multiGPU}
\end{figure}

\begin{figure}[t]
\centering
\includegraphics[width=0.85\linewidth]{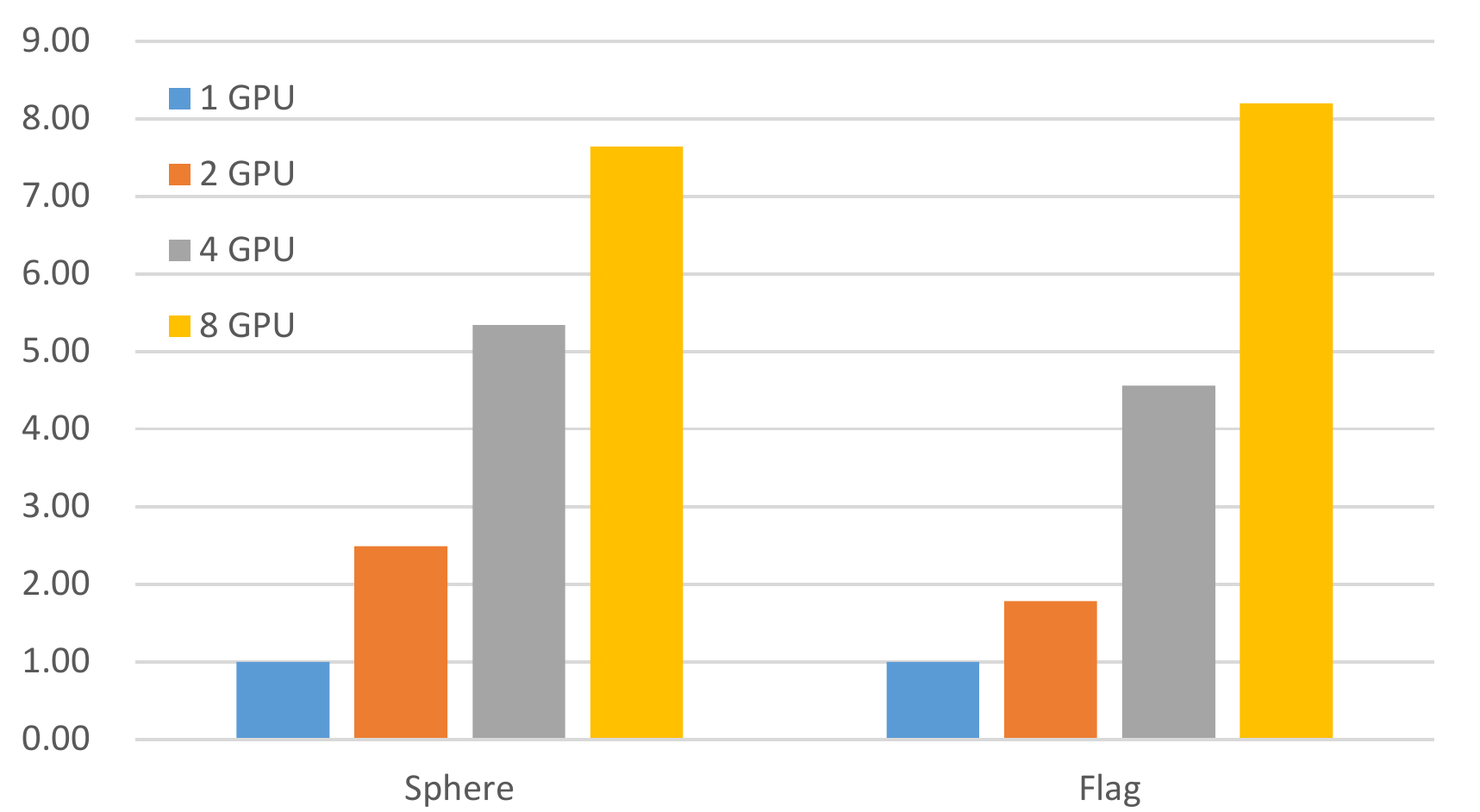}
\caption{ {\bf Scalability with the Number of GPUs:} This figure highlights the speedups achieved with 2, 4, and 8 GPUs, respectively, on benchmarks Flag and Sphere. The performance data are collected on the workstation with 8 NVIDIA Titan V GPUs. The quasi-linear speedups demonstrate good scalability of our GPU algorithm with the number of GPUs.}
\label{fig:speedup-8gpu}
\end{figure}

\begin{figure}[t]
\centering
\includegraphics[width=0.9\linewidth]{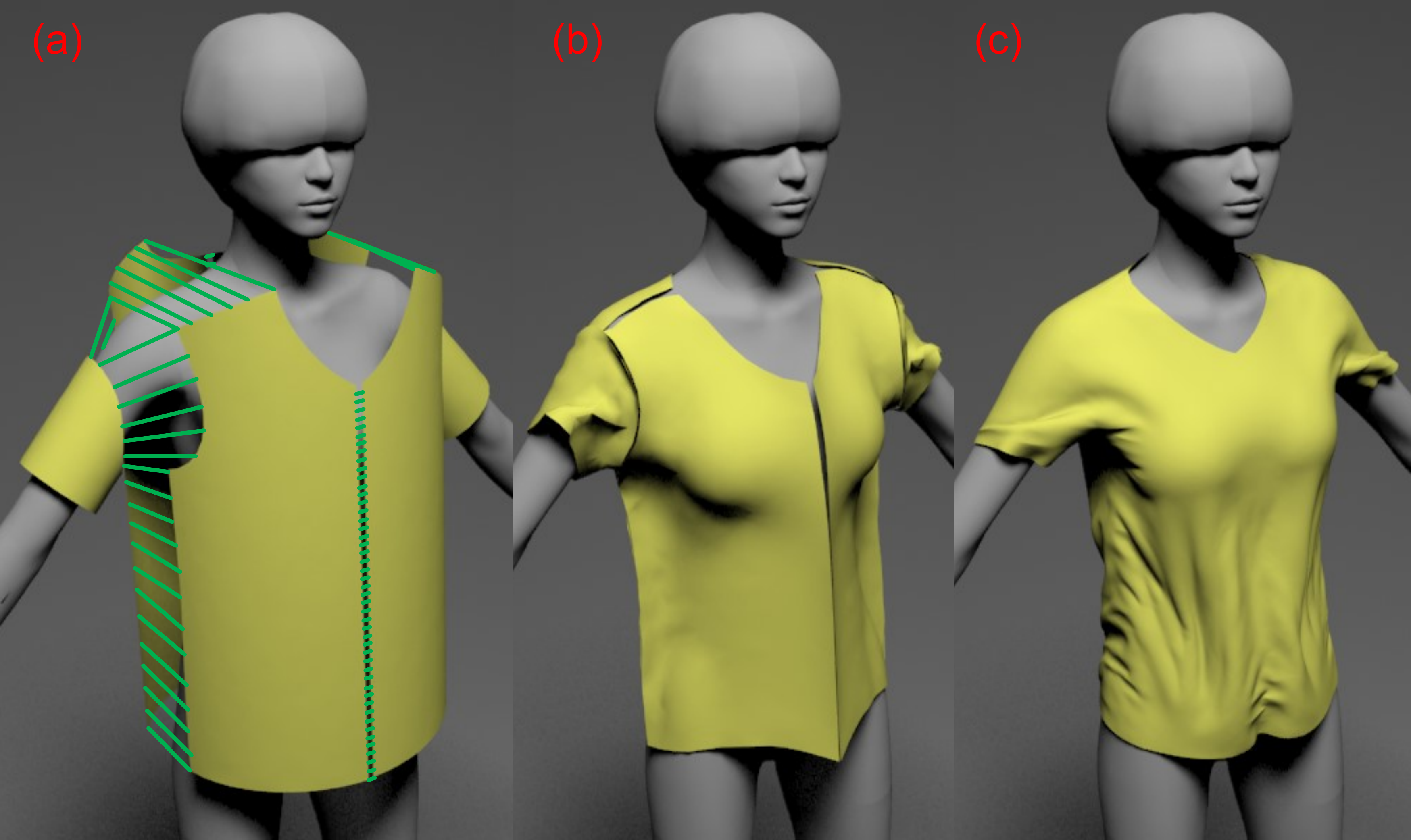}
\caption{{\bf Interactive Stitching:} We use our interactive simulation algorithm to stitch  cloth pieces: (left) A designer first assigns stitching pairs interactively using a simple mesh with $5K$ triangles; 
(middle) P-Cloth can perform the stitching operation at 10+ fps on 4 GPUs; (right) final result  obtained after stitching with $316$K mesh. The stitching is first performed with the coarse mesh ($5K$ triangles) and refined to the detailed mesh ($316K$ triangles).
}
\label{fig:stitching}
\end{figure}

\noindent {\bf Memory Overhead:}
We need to store the original mesh, matrices generated for implicit integration as well as the acceleration data structures used for faster collision detection. Most prior algorithms for collision detection are based on spatial hashing~\cite{i3d18} and BVHs~\cite{cama16}. In practice, these acceleration data structures can have a high memory overhead, especially for techniques based on BVHs. Not only do we need to store the hierarchy, we also need to store the BVTT front and all possible triangle or feature pairs to perform the exact elementary CCD tests, including VF (vertex-face) or EE (edge-edge) continuous tests. 
In contrast, techniques based on spatial hashing have a lower memory overhead, as we don't need to store the front. 

The memory overhead of P-Cloth is dominated by collision detection data structures and pairwise elementary tests.
The collision detection algorithm uses considerable memory to perform the pairwise triangle/VF/EE tests in parallel. We show the average memory overhead over all the frames in Fig.~\ref{fig:memory-ratio}.

We also highlight the memory utilization as a function of number of GPUs. As the number of GPUs increase, we observe almost linear reduction in the memory overhead. This makes it possible to perform interactive simulation on very complex meshes.
For the Sphere-1M benchmark, the memory overhead can be as high as $29$ GB for some close proximity configurations with a very high number of potential pairs. Current GPUs may not have sufficient memory, so we need a 4-GPU system to achieve almost interactive performance. If we use 1-GPU or a 2-GPU system, the spatial data structures and collision pairs will not fit in the GPU memory, and the resulting system would not run or be too slow.
We highlight the  memory overhead for the  Sphere-1M benchmark by varying the number of GPUs
in Figure~\ref{fig:multiGPU}. 

\noindent {\bf Scalability:}
Figure~\ref{fig:speedup-8gpu} highlights the scalability of our GPU-based parallel algorithm with respect to the number of GPUs. We collect the performance data on a workstation with 8 NVIDIA Titan V GPUs, and obverse quasi-linear speedups on Benchmark Sphere and Flag ($7.64$X and $8.23$X with 8 GPUs, respectively).

\noindent {\bf Interactive Stitching:}
Cloth piece stitching is an essential tools for 3D garment design~\cite{Zhang05}. As the result of the high performance of the P-Cloth system, our simulator can perform the stitching at interactive rate (the video).  
The stitching is first performed at a coarse level ($5K$ triangles), then the cloth is refined to a detailed mesh ($316K$ triangles) to improve the simulation fidelity in terms of wrinkles and folds. 
As shown in Fig.~\ref{fig:stitching} and the video, the T-shirt model is stitching together using several cut pieces. The entire process (both the coarse and the refined levels) can be simulated at $10+$ fps on the 4-GPU workstation.


\section{Comparison and Analysis}
\label{sec:compare}

In this section, we compare the features and performance of our approach with prior parallel cloth simulation algorithms.

\subsection{Comparison}

Compared with prior GPU-based cloth simulation systems~\cite{cama16,tang-siga18}, the main benefits of our algorithm include much lower memory overhead due to the parallel computation (Figure~\ref{fig:memory-ratio}), improved runtime performance (Table~\ref{tbl:performance}), {and equal simulation quality (video).} The reduced memory overhead and smaller working set on each GPU for P-Cloth can considerably improve the runtime performance and can also result in super-linear speedups on some benchmarks. With the enhanced spatial hashing data structures (see Fig.~\ref{fig:sh-pipeline}), P-Cloth can perform self-collision culling and perform fewer elementary tests for CCD. This suggests that the memory overhead and the working set sizes can have considerable impact on the performance of collision detection and cloth simulation algorithms.



\subsubsection{Matrix-free Methods and Parallel Cloth Simulation}
{Some researchers have used matrix-free approaches for time integration to reduce the memory overhead~\cite{Eike13,PrabhuneS20}. However, these approaches tend to be useful for large-scaled simulations (i.e., node number $> 10M$) and they trade-off memory overhead with the computational costs. Such methods are not frequently  used for medium-sized simulations (i.e., node number $\in [500K, 10M]$). }

{Compared with explicit matrix storage and assembly, matrix-free methods have the benefit of reduced memory overhead. However, our matrix assembly and PCG solver takes less  than $100$ MB GPU memory, which is about $15\%$ of the $8$--$9$ GB memory used for collision handing. As shown in Fig.~\ref{fig:memory-ratio} and Fig.~\ref{fig:multiGPU}, most of the memory overhead arises from collision handling due to the large number of potentially colliding triangle/VF/EE pairs.
While we can use matrix-free methods to reduce the memory overhead of implicit time integration, it would increase the computational overhead to compute the matrix elements on-the-fly. Selle et al.~\shortcite{Selle09} use a matrix-free approach to simulate clothes with several million vertices using MPI over a cluster. However, it takes several or tens of minutes per frame on workstations with 8-16 CPU processors, with $45-60\%$ running time for time integration.
Given that the matrix storage and matrix assembly are not the bottlenecks in our approach, it is not clear that matrix-free methods would offer improved performance on multi-GPU systems.
}

\subsection{High-resolution Cloth Meshes}
We highlight the average running time of our algorithm on complex  benchmarks with $510 - 1650K$ triangles (see  Table~\ref{tbl:performance}). The actual performance, including frame rate and memory overhead, also depends on the relative configuration and number of triangle pairs in close proximity.  P-Cloth takes about $200-500$ms per frame on the 4-GPU system, which includes a high number of wrinkles, folds and self-collisions (see Fig.~\ref{fig:resolution-compare} and the video).

Comparing to the parallel cloth simulation algorithms on a CPU cluster~\cite{Selle09,Ni15,Liang18} or a hybrid  combination of CPU and GPU~\cite{PKS10}, our multi-GPU based algorithm is about $10$X faster on performance for complex benchmarks.
Recent cloth simulation algorithm proposed by Jiang et al.~\shortcite{Jiang2017} takes about 2 minutes per frame for complex benchmarks with $1.4-1.8$M triangles on an Intel Xeon system with multiple CPU cores.  Not only are the underlying processors used to run these cloth simulation systems different, but also the techniques for collision detection and response computation in ~\cite{Jiang2017} are different from P-Cloth. That makes it hard to make a fair comparison with prior methods. To the best of our knowledge, P-Cloth is the first cloth simulation algorithm that can perform almost interactive cloth simulation on complex meshes using commodity workstations.



\section{Conclusion and Limitations}
\label{sec:conclude}

We present a multi-GPU based cloth simulation algorithm, P-Cloth,  for high resolution meshes. It is based on three novel multi-GPU algorithms: SpMV for time integration, matrix assembly, and collision handling. Our approach is designed for sparse linear systems with a dynamic layout, which are widely used for robust cloth simulation. We have evaluated the performance on complex cloth meshes with more than a million triangles and observe almost linear speedups on workstations with 4 or 8 GPUs. P-Cloth is the first interactive cloth simulation algorithm that can handle complex cloth meshes on commodity workstations.


Our approach has several limitations. For cloth configurations with folds or self-collisions, collision detection remains a bottleneck. 
{Although our Pipelined SpMV algorithm is a general solution for all distributed systems, our work queue and data transfer algorithms are limited to fat-tree topologies for GPU interconnection and assume that the topology is known apriori. It will be useful to evaluate the performance on other topologies.}
The overall speedup can vary depending on the cloth configuration and data synchronization overhead.
The robustness of the system is governed by contact force computation and a non-linear impact zone solver for penetration handling. Our current parallel algorithm is based on implicit time integrator, and other methods based on 
 projective dynamics, Anderson acceleration, or ADMM may offer better stability.

{Our cloth simulator can be used not only for VFX application, but also is applicable for CAD and gaming/VR applications.  Our parallel (and almost interactive) simulation can be used for cloud-based gaming, that uses remote servers and streams the results to the device. The parallel SpMV algorithm is also useful for other scientific applications.}

There are many avenues for future research. In addition to overcoming the limitations, we  need to evaluate the scalability of our approach on workstations with higher number of  GPUs and different interconnect topologies.  It may be possible to improve the performance by exploiting the memory hierarchy and cache layouts of modern GPUs. Our novel parallel algorithms for SpMV, implicit integration and collision detection can also be used to accelerate other simulations on multi-GPU systems. In particular, SpMV is a fundamental computation for many GPU-based scientific applications~\cite{filippone2017sparse} and it may be useful to apply our Pipelined SpMV algorithm (Section 3.1) and matrix assembly (Section 3.2)  to other applications like finite-element simulation (FEM), material point method (MPM), etc.



\bibliographystyle{ACM-Reference-Format}
\bibliography{main}

\end{document}